\documentclass[sigconf]{acmart}
\AtBeginDocument{%
  \providecommand\BibTeX{{%
    \normalfont B\kern-0.5em{\scshape i\kern-0.25em b}\kern-0.8em\TeX}}}

\newcommand{\eg}{\textit{e.g.}}
\newcommand{\ie}{\textit{i.e.}}
\newcommand{\cf}{\textit{c.f.}}

\copyrightyear{2022}
\acmYear{2022}
\setcopyright{rightsretained}
\acmConference[CHI '22]{CHI Conference on Human Factors in Computing Systems}{April 29-May 5, 2022}{New Orleans, LA, USA}
\acmBooktitle{CHI Conference on Human Factors in Computing Systems (CHI '22), April 29-May 5, 2022, New Orleans, LA, USA}
\acmDOI{10.1145/3491102.3517457}
\acmISBN{978-1-4503-9157-3/22/04}

\usepackage{booktabs} 
\usepackage{stfloats}
\usepackage{soul} 
\usepackage{multirow}
\usepackage{array}
\usepackage{color}
\usepackage{colortbl}
\usepackage{subcaption}
\usepackage{enumitem}

\usepackage{bm} 

\newcommand{\thicktablehline}{\specialrule{0.75pt}{0pt}{0pt}}

\newcolumntype{P}[1]{>{\centering\arraybackslash}p{#1}}
\newcolumntype{C}[1]{>{\centering\arraybackslash}p{#1}}

\definecolor{youngho}{RGB}{27,158,119}
\definecolor{bongshin}{RGB}{217,95,2}
\definecolor{eunkyoung}{RGB}{102,166,30}
\definecolor{diana}{RGB}{20, 201, 192}

\definecolor{revisedcolor}{RGB}{0,0,255}

\newcommand{\revised}[1]{#1}

\newcommand{\cameraready}[1]{#1}

\definecolor{tableheader}{HTML}{EFEFEF}
\definecolor{tablegrayline}{HTML}{d0d0d0}

\newcolumntype{R}[1]{>{\RaggedLeft\arraybackslash}p{#1}}

\newcommand{\symbolmic}{\raisebox{-1.5pt}{\includegraphics[width=10pt]{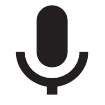}} }
\newcommand{\symboldelete}{\raisebox{-1.5pt}{\includegraphics[width=10pt]{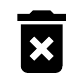}} }

\newcommand{\ipstart}[1]{\vspace{1mm} \noindent{\textbf{\textit{#1.}}}}

\newcommand{\autorefsuffix}[2]{\hyperref[#1]{\autoref{#1}#2}}

\newcommand{\circledigit}[1]{\textbf{\normalsize{\textsf{\textcircled{\footnotesize{#1}}}}}}

\definecolor{patternactivity}{HTML}{f5e9d0}
\definecolor{patternactivitytext}{HTML}{bd7004}

\definecolor{patterntime}{HTML}{ebc0ce}
\definecolor{patterntimetext}{HTML}{d1567f}

\definecolor{patterneffort}{HTML}{d5eef2}
\definecolor{patternefforttext}{HTML}{2ebbd1}

\definecolor{patterncondition}{HTML}{f5e9d0}
\definecolor{patternconditiontext}{HTML}{bd7004}

\definecolor{colorredbg}{HTML}{fcb1b1}
\definecolor{colororangebg}{HTML}{ffd49c}

\definecolor{colorbluebg}{HTML}{a3d8e5}

\definecolor{indirect}{HTML}{757575}

\newcommand{\labelphantom}[1]{%
  \parbox{0pt}{\phantomsubcaption\label{#1}}%
}

\acmSubmissionID{1008}

\begin{document}

\title{MyMove: Facilitating Older Adults to Collect In-Situ Activity Labels on a Smartwatch with Speech}


    \author{Young-Ho Kim}
    \affiliation{%
          \institution{University of Maryland}
          \city{College Park}
          \state{MD}
          \country{USA}
    }
    \email{yghokim@younghokim.net}

    \author{Diana Chou}
    \affiliation{%
          \institution{University of Maryland}
          \city{College Park}
          \state{MD}
          \country{USA}
    }
    \email{dchou4@umd.edu}
    
    \author{Bongshin Lee}
    \affiliation{%
          \institution{Microsoft Research}
          \city{Redmond}
          \state{WA}
          \country{USA}
    }
    \email{bongshin@microsoft.com}
    
    \author{Margaret Danilovich}
    \affiliation{%
          \institution{CJE SeniorLife}
          \city{Chicago}
          \state{IL}
          \country{USA}
    }
    \email{margaret-wente@northwestern.edu}
    
    \author{Amanda Lazar}
    \affiliation{%
        \institution{University of Maryland}
        \city{College Park}
        \state{MD}
        \country{USA}
    }
    \email{lazar@umd.edu}
    
    \author{David E. Conroy}
    \affiliation{%
          \institution{Pennsylvania State University}
          \city{University Park}
          \state{PA}
          \country{USA}
    }
    \email{conroy@psu.edu}

    \author{Hernisa Kacorri}
    \affiliation{%
        \institution{University of Maryland}
        \city{College Park}
        \state{MD}
        \country{USA}
    }
    \email{hernisa@umd.edu}
    
    \author{Eun Kyoung Choe}
    \affiliation{%
        \institution{University of Maryland}
        \city{College Park}
        \state{MD}
        \country{USA}
    }
    \email{choe@umd.edu}

\settopmatter{authorsperrow=4} 

\renewcommand{\shortauthors}{Kim, Y.-H., Chou, D., Lee, B., Danilovich, M., Lazar, A., Conroy, D.E., Kacorri, H., and Choe, E.K.}

\begin{abstract}
Current activity tracking technologies are largely trained on younger adults' data, which can lead to solutions that are not well-suited for older adults. 
To build activity trackers for older adults, it is crucial to collect training data with them. 
To this end, we examine the feasibility and challenges with older adults in collecting activity labels by leveraging speech.  
Specifically, we built MyMove, a speech-based smartwatch app to facilitate the in-situ labeling with a low capture burden.
We conducted a 7-day deployment study, where 13 older adults collected their activity labels and smartwatch sensor data, while wearing a thigh-worn activity monitor.
Participants were highly engaged, capturing 1,224 verbal reports in total. 
We extracted 1,885 activities with corresponding effort level and timespan, and examined the usefulness of these reports as activity labels.
We discuss the implications of our approach and the collected dataset in supporting older adults through personalized activity tracking technologies.



\end{abstract}

\ccsdesc[500]{Human-centered computing~Empirical studies in ubiquitous and mobile computing}
\ccsdesc[500]{Human-centered computing~Sound-based input / output}

\keywords{activity labeling, older adults, smartwatch, speech interaction, experience sampling method}


\begin{teaserfigure}
    \includegraphics[width=\textwidth]{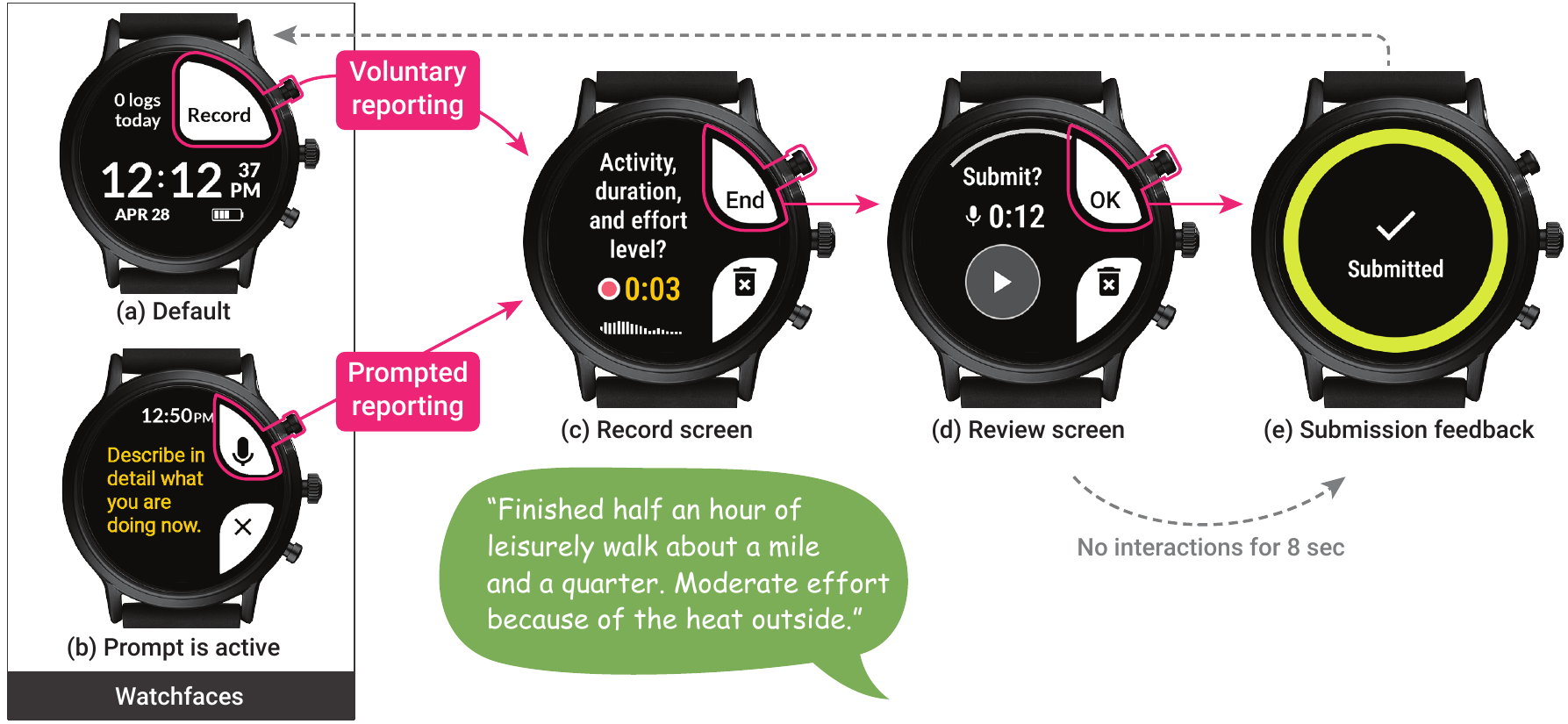}
    \labelphantom{fig:design:flow:watchface:default}
    \labelphantom{fig:design:flow:watchface:prompt}
    \labelphantom{fig:design:flow:record}
    \labelphantom{fig:design:flow:review}
    \labelphantom{fig:design:flow:feedback}
    \vspace{-4mm}
    \caption{MyMove supports collecting in-situ activity labels using speech on a smartwatch. People can initiate the reporting from the watchface either voluntarily (a) or upon a prompt message (b); describe an activity, time span, and effort level (c); and review \& submit the recording (d). MyMove displays a visual confirmation after the submission (e). The example verbal report is from P7. \cameraready{Please refer to our supplementary video which demonstrates the interactions.}}
    \label{fig:design:flow}
    \Description{An interaction flow of MyMove, demonstrating the two reporting methods, voluntary and prompted reporting. An example verbal report quoted here is ``Finished half an hour of leisurely walk about a mile and a quarter. Moderate effort because of the heat outside.'' This flow is described in detail in Section \ref{sec:datacollection}.}
\end{teaserfigure}

\maketitle

\section{Introduction}

Scarcity of older adults' activity datasets may lead to biased and inaccurate activity recognition systems. For example, a recent study showed that Fitbit Ultra, a consumer health tracking device, significantly under-reports steps at slow speed of 0.9 m/s, a representative walking speed of older adults~\cite{wong2018bit}. When people walk slowly, with a cane, or a walker, such activity recognition systems have a tendency to not register steps accurately. 
A recent study looking at older adults' technology usage for activity tracking shows that more than a half do not trust the accuracy of these devices~\cite{aarp2016}, which are typically trained on younger adults data. To develop activity tracking systems that are inclusive of and beneficial to older adults, it is imperative to collect older adults' movements and activity data.

Activity tracking technologies can provide meaningful feedback that supports people's motivations, playing an important role in enhancing physical activity~\cite{Vargemidis2020WearableOlderAdults, french2014behaviour, mercer2016behavior}. Like individuals in many age groups, physical activity is important for older adults, favorably influencing \revised{their healthy daily routine~\cite{Fan2012ConsiderationOlderAdults} and active life expectancy~\cite{chodzko2009exercise}, chronic health conditions including coronary heart disease, hypertension, and type 2 diabetes~\cite{warburton2006health}, psychological health and wellbeing~\cite{chodzko2009exercise}, enjoyment~\cite{phoenix2013narratives, phoenix2014pleasure}, and social wellbeing~\cite{baltes1990successful}}. 
However, the adoption rate of activity tracking technologies for older adults is relatively low (\eg, 10\% for age 55+ whereas 28\% for ages 18--34 and 22\% for ages 35--54~\cite{gallup2019}). Meanwhile, studies continuously report that younger, more affluent, healthier, and more educated groups are more likely to use activity tracking technologies~\cite{gallup2019,Chandrasekaran2020WearableSurvey,pew2019, macridis2018consumer}.

We suspect that the current activity tracking technologies are designed with little understanding of older adults' lifestyles and perspectives (\eg, types of activities they engage in and care about) \revised{and do not account for heterogeneous} physiological characteristics (\eg, gait and locomotion~\cite{Boss1981AgeRelated}). Our ultimate goal is to support older adults\revised{' agency} by designing and developing personalized activity tracking technologies that better match their preferences and patterns. As a first step, we set out to develop an activity labeling tool that older adults \revised{can use to} collect \textit{in-situ} activity labels along with their sensor data. These labels could be used to train and fine-tune classifiers based on inertial sensors.

To this end, we conducted a 7-day deployment study with 13 older adult participants (age range: 61--90; average: 71.08), where they collected activity \revised{descriptions} while wearing a smartwatch and a thigh-worn activity monitor; the thigh-worn activity monitor served as a means for collecting ground-truth sensor data for our analysis \revised{and later model development}.
To facilitate collecting in-situ \revised{descriptions} with a low data capture burden, we designed and developed an Android Wear reporting app, called \textbf{MyMove}, leveraging speech input, \revised{an accessible modality for many older adults~\cite{pradhan2020use}. With MyMove on a smartwatch, participants can describe} activity type, associated timespan, and perceived effort level. Many smartwatches are equipped with a microphone, which allows people to flexibly describe their activities using speech. As an on-body device, a smartwatch can collect continuous activity sensing data and deliver notifications, which is necessary to collect in-situ data through an experience sampling method (ESM)~\cite{larson2014experience}. \revised{Furthermore, prior work co-designing wearable activity trackers with older adults showed that the ``watch-like'' form factor was mostly preferred due to its ability to tell time, on-body position, and public acceptance~\cite{Vargemidis2021IrrelevantGadgets}.} 
Through our deployment study, with a focus on feasibility, we explore the following questions: (1) How do older adults capture their activities using speech on a smartwatch? and (2) How useful are their verbal reports as an information source for activity labeling?

Our results show that participants were highly engaged in the data collection process, submitting a total of 1,224 verbal reports (avg. 13.45 reports per day per participant) and wearing the smartwatch and monitor throughout the seven-day study period. 
From these reports, we extracted 1,885 activities with 29 different activity types that comprehensively capture participants' daily lifestyles. 
Participants provided time-related information for about a half of the activities but they were more likely to provide complete time information when reporting a single activity or when reporting voluntarily as opposed to being prompted.
Participants' effort level categories were aligned with sensor-based intensity metrics in the corresponding time segments. However, activities that participants evaluated as moderate to high intensity did not meet the standard intensity level according to the sensor-based intensity measurements. 
All of the 1,224 verbal reports were valid and could be transcribed and understood by a researcher. Furthermore, the word error rates of these reports by two state-of-the-art speech recognition systems were relatively low: 4.93\% with Microsoft Cognitive Speech and 8.50\% with Google Cloud Speech.
Through our study, we demonstrated that by leveraging speech, MyMove can facilitate collecting useful activity labels. We also identified how we can further improve speech-based activity labeling tools for older adults; for example, by leveraging multi-device environments to collect more accurate and fine-grained data and by providing self-monitoring feedback to enhance engagement.
The key contributions of this work are: 
\begin{enumerate}[leftmargin=*]
    \item Design and development of MyMove, an Android Wear reporting app for supporting older adults in collecting their activity \revised{descriptions} with a low data capture burden by leveraging speech input on a smartwatch. 
    \item Empirical results from a deployment study conducted with 13 older adults using MyMove, demonstrating the \textit{feasibility} of collecting rich in-situ activity \revised{descriptions} from older adults via speech.
    \item Examining the characteristics and \textit{usefulness} of the data collected with MyMove, in terms of \revised{activity type,} time, and effort level as well as the quality of the voice recording (automatic speech recognition error). 
\end{enumerate}

\section{Related Work}
In this section, we cover the related work in the areas of (1) understanding older adults' activities, (2) collecting in-situ behavioral data, and (3) in-situ data labeling for human activity recognition.

\subsection{Understanding Older Adults' Activities}
Researchers, healthcare providers, and government officials have been interested in understanding daily activities of older adults because it helps establish and improve health-related guidelines, policies, and interventions~\cite{Gerling2020CriticalReflections, Sparling2015Recommendations, Brawley2003PrommotingPA}. 
Researchers have defined ``activity'' differently depending on their research focus. 
For example, there is a focus on assessing the independence/dependence with functional tasks, as reflected in the concept of \textit{ADL} (Activities of Daily Living---basic self-maintenance activities, such as eating, dressing, bathing, or toileting)~\cite{Katz1963ADL} and \textit{IADL} (Instrumental ADL---higher-level activities that require complex skills and mental load, such as making a phone call, shopping, housekeeping, or financing)~\cite{Lawton1969IADL}.
Another subset of research categorizes activities based on the level of energy expenditure (\cf, classification of energy costs of daily activities~\cite{Ainsworth1993Compendium}), as reflected in many physical activity questionnaires they developed to assess older adults' intensity-specific duration for behavior (\eg, MOST~\cite{Gardiner2011MOST}, CHAMPS~\cite{Stewart2001CHAMPS}, LASA~\cite{Visser2013LASA}).

Domestic and leisure activities are prevalent in older adults' daily activities~\cite{Kan2021HowDoOlderAdults, McKenna2007WhatOlderPeopleDo, Moss1982TimeBudgetOlder, Horgas1998OlderAdultsLife}. According to the national time use surveys from 14 countries, older adults (aged 60--75) spent around 6 hours on leisure and $\geq$2.5 hours on domestic work daily~\cite{Kan2021HowDoOlderAdults}.
From the interviews with U.S. older adults, Moss and Lawton found that participants spend about 5 hours a day on obligatory personal care \& household activities and more than 6 hours a day on discretionary leisure activities~\cite{Moss1982TimeBudgetOlder}. Another study with Australian older adults reported that participants spend the longest time on solitary leisure (\textit{avg.} 4.5 hours a day) excluding sleep, followed by IADL (\textit{avg.} 3.1 hours a day), social leisure (\textit{avg.} 2.7 hours a day), and ADL (\textit{avg.} 2.6 hours a day)~\cite{McKenna2007WhatOlderPeopleDo}. 

Researchers have further examined what kinds of activities older adults engage in during their leisure time\revised{~\cite{Kan2021HowDoOlderAdults} grouping them as \textit{active}  (\eg, relaxing, socializing, volunteering, organization work, religion, going out, sports and exercising) and \textit{passive}  (\eg, reading, listening to the radio, watching television, and browsing the internet on a computer) with the latter often involving screen time.} \textit{Screen time} is one of the most prevalent leisure time activities~\cite{ONeill2016SedenteryTypes}; studies consistently report that older adults spend longer than 2 hours a day watching TV (\eg, \textit{avg.} 2.5 hours~\cite{Horgas1998OlderAdultsLife}, \textit{avg.} 3.5 hours~\cite{Moss1982TimeBudgetOlder}, and over 3 hours for 54.6\% of an older population~\cite{Harvey2013PrevalenceSedentary}).
Screen time is known to be a strong indicator of discretionary sedentary behaviors (\ie, low energy expenditure activities in a seated or reclined posture while awake~\cite{SBRN2012}). Decreased physical activity during leisure time and increased sedentary time is another common characteristic of older adults \revised{that may be disproportionately affected by many other factors such as the socioeconomic status of their neighborhood~\cite{annear2009leisure}. } The U.S. national surveys in 2015--2016 revealed that 64\% of older adults aged 65+ reported being inactive (\ie, no moderate or vigorous-intensity activity for 10 minutes per day), and 53\% reported that they sit longer than 6 hours a day~\cite{Wiebe2018SittingTime}. In a similar vein, a study using an accelerometer sensor (ActiGraph) found that older adults aged 70+ in the urban UK spend less than 30 minutes on moderate-to-vigorous physical activities and the duration significantly drops with age~\cite{Davis2011ObjectivelyMeasuredPA}.

\cameraready{This body of knowledge---that is typically based on retrospective recall, surveys, and automated sensing---provides a general understanding of older adults' activities and time use. In our work, however, the purpose of collecting older adults' activities is quite different: going beyond understanding how older adults spend their time, we aim to examine the feasibility of \textit{creating a training dataset} that contains older adults' activity 
patterns. To this end, we employ a low-burden, in-situ data collection method that older adults can partake in to collect fine-grained data of their activities.}  

\subsection{Collecting In-Situ Behavioral Data}

Methods that rely on retrospective recall, such as interviews or surveys, are subject to
recall bias~\cite{Harari2016SmartphoneForBehavioralScience}, which may be affected by the nature of an event and people's experiences. For example, in responding to a survey, people were likely to accurately estimate the past duration of intensive physical activities~\cite{Bonnefoy2001SimultaneousValidation, Jacobs1993SimultaneousEvaluation, Schrack2016AssessingPASensors}, whereas they were likely to underestimate or omit light and sedentary activities~\cite{Sallis2000AssessmentPA, Jacobs1993SimultaneousEvaluation, Bonnefoy2001SimultaneousValidation, Lee2014UsingAccelerometers, Schrack2016AssessingPASensors}.
To collect more ecologically valid self-report data, researchers devised Diary Study~\cite{Bolger2003DiaryMethods} and Experience Sampling Method (ESM, often interchangeable with ecological momentary assessment or EMA)~\cite{larson2014experience}. 
Both methods have been employed before the widespread use of smartphones, but smartphones and their notification capability have made it much easier to facilitate these methods.
In Diary Studies, people are expected to capture self-report data once (or more) a day using pen and paper or diary apps. Although Diary Studies help researchers collect in-situ self-report data, there can be a delay between when an event happens and when that event is captured. To further reduce recall bias, ESM employs notifications (defined by a certain prompting rule) to signal when to capture data, and people are expected to capture data at the moment of (or shortly after) receiving the notification. Researchers typically employ ESM to collect \textit{brief} self-report data \textit{frequently}. Therefore, in an ESM study, it is important to strike the balance between researchers' data collection needs and participants' data capture burdens.


To reduce data capture burdens, researchers have explored smartwatches as a new means to facilitate ESM~\cite{Yan2020LightweightInSituSmartwatch, Hernandez2016WearableESM};
wearing smartwatches allows for high awareness of and alertness to incoming notifications with glanceable feedback~\cite{Cecchinato2017UXSmartwatch, Pizza2016SmartwatchInVivo}. 
In terms of the notification delivery, prior work has demonstrated that smartwatch-based ESM can yield shorter response delays~\cite{Hernandez2016WearableESM}, higher response rates, and EMA experiences perceived as less distracting~\cite{Intille2016microEMA, Ponnada2017microEMAResponseRate} when compared to smartphones. 
On the other hand, an inherent drawback of smartwatches for ESM is their small form factor, which can make it laborious to enter data. \revised{Thus, approaches typically employed on smartphones (\eg, entering data via a text box) are inefficient.}
 To ease the data entry, researchers have explored more effective input methods such as the ROAMM~\cite{Kheirkhahan2019ROAMM} and PROMPT~\cite{Manini2019OlderAdultsSmartwatch} frameworks, which support radial scales and bezel rotation to specify pain level and activity type. 
Others have combined touch and motion gestures for answering Likert scale questions~\cite{Yan2020LightweightInSituSmartwatch}. 

These prior studies predominantly incorporated graphical widgets with touch/hand gestures for structured questions with simple choices (\eg, ``yes'' or ``no''). One input modality on a smartwatch that has not been actively considered for ESM on a smartwatch is \textit{speech}, which is widely embedded in consumer devices and digital systems~\cite{clark2019state}.
\revised{When people speak, they tend to be faster~\cite{Ruan2018ComparingSpeechKeyboard} and more expressive~\cite{chalfonte1991expressive,revilla2020testing} than when they type.} 
Speech input requires little to no screen space and researchers found that speech commands can be easier to perform than using graphical widgets on mobile devices (\eg,~\cite{Srinivasan2020InChorus, Kim2021Data@Hand}).
Recent work has shown promise for speech input for in-situ data collection on digital devices (\eg, exercise logging on a smart speaker~\cite{Luo2020TandemTrack}, food journaling on a smartphone~\cite{Luo2021FoodScrap}). For example, Luo and colleagues deployed a speech-based mobile food journal and found that participants provided detailed and elaborate information on their food decisions and meal contexts, with a low perceived capture burden~\cite{Luo2021FoodScrap}.
Using speech input on a smartwatch poses great potential for lowering the data capture burden while enhancing response rate in EMA studies. It allows us to \revised{mitigate} touch interactions that involve on-screen finger movement, such as scrolling, which may be burdensome for older adults~\cite{Bakaev2008FittsLawOlderAdults}. \revised{Given that voice-based interfaces tend to be accessible for many older adults~\cite{pradhan2020use} (including those with low technology experience),} in this paper, we explore how older adults leverage speech input on a smartwatch to collect in-situ activity data in an open-ended format. \revised{This} is a novel approach to prior ESM studies that collected responses to structured questions (\eg, multiple choice, Likert scale).

\subsection{In-Situ Data Labeling for Human Activity Recognition}
Another relevant topic to our work is Human Activity Recognition (HAR), an automated process of relating sensor stream time segments with various human activities (\eg, walking, running, sleeping, eating)~\cite{Lara2013HARSurvey}. HAR has been extensively applied to a wide range of technologies, from \revised{broader} consumer fitness trackers to \revised{specialized} tracking systems for older adults in capturing physical activities and ADLs or detecting falling or frailty~\cite{Gerling2020CriticalReflections, Vargemidis2020WearableOlderAdults, Vargemidis2021IrrelevantGadgets}. The quality of an HAR model depends on how sensor data (\ie, input) were collected~\cite{Lara2013HARSurvey}; models trained with the sensor data captured in the lab tend to yield less accuracy when tested outside~\cite{Foerster1999Detection}. However, gathering both ground-truth activity labels (\ie, the type of activity, the start and end time of an activity) and sensor data in the natural context of daily life is generally challenging because it may not be ethical or feasible for researchers to observe participants' activity outside the lab~\cite{Hoque2014VocalDiary}.

To enable in-situ collection of both the sensor data and the activity labels, the UbiComp and HCI communities have proposed mobile and wearable systems that allow participants to label their own activities, while collecting sensor data in the background (\eg, \cite{Samyoun2021VoiSense, Vaizman2018ExtraSensoryApp, Mondol2018WaDa}). For example, VoiSense~\cite{Samyoun2021VoiSense} is a conversational agent on Apple Watch that allows people to capture the physiological or motion sensor data for a designated duration and then specify a label for the session, \revised{though, it has not been evaluated yet with users.} 
ExtraSensory App~\cite{Vaizman2018ExtraSensoryApp} is an in-situ activity labeling system that consists of a mobile and smartwatch app. On the mobile app, people can review their activity history and labels for past or near-future time segments. The smartwatch app complements the mobile app by forwarding notifications or receiving binary confirmations about the current status (\eg, ``\textit{In the past 2 minutes were you still sitting?}''). When labeling on the mobile app, people can select multiple labels from a predefined list (\eg, \textit{Sitting} + \textit{At work}) that best describes the time segment. \revised{Data collected with the ExtraSensory App typically include younger adults (ages 18-42).} 

Our work extends this line of research on collecting in-situ activity labels in two ways. First, unlike prior systems that primarily target younger adults, we aim to work with older adults with interfaces that are specifically designed for this population (see Design Rationale \hyperref[sec:dr1]{DR1} in Section ~\ref{sec:dr}). Second, unlike VoiSense and ExtraSensory App, which collect structured label data through multiple steps of speech or touch inputs, we collect activity information as an unstructured verbal description on the activity type, associated timespan, and perceived level of effort. 
In doing so, we explore how useful such utterances are as a source of information for activity labeling and discuss the implications of our findings for how to design low-burden in-situ activity labeling systems suitable for older adults.

\section{MyMove}
As a low-burden activity reporting tool \revised{intended} for older adults, we designed and developed MyMove~(\autoref{fig:design:flow}), a speech-based Android Wear app.
MyMove allows people to submit a verbal description of their activities (which we call a verbal \textit{report} throughout the paper) in two different methods: \revised{(i) report \textit{voluntarily} at any time or (ii) report when they are \textit{prompted} by ESM notifications.}
MyMove asks people to include \textbf{activity}, \textbf{time/duration}, and \textbf{perceived effort level} in their verbal report. The activity and associated timespan are the two essential components \revised{of} labeling sensor data for Human Activity Recognition~\cite{Lara2013HARSurvey}: \revised{activity labels can be extracted from activity descriptions} and the timespan \revised{connects the activity and} the sensor values.
Capturing the \textit{perceived} level of effort is important because it varies from person to person even when they perform the same activity (\eg, the number of repetitions, speed, weight lifted)~\cite{Borg1990Scaling}.
In the background, MyMove captures sensor data streams and transmits them to a backend server.
In this section, we describe our design rationales and the MyMove system along with the implementation details.

\subsection{Design Rationales}
\label{sec:dr}

\phantomsection{}
\label{sec:dr1}
\ipstart{DR1: Prioritize Older Adults} \revised{Both the form factor and interaction modalities of MyMove are informed by prior work with older adults in support of smartwatches in the context of activity tracking (\eg,~\cite{Vargemidis2021IrrelevantGadgets, fernandez2017my}), voice as an accessible input modality for many older adults~\cite{pradhan2020use}, and large target buttons associated with tapping or pressing~\cite{Carmeli2003AgingHand, Motti2013OlderAdultsInputTechnique}.}

\revised{We carefully selected hardware (i.e., Fossil Gen 5) that has a relatively big display among other smartwatch options with similar sensing.} 
Interacting with Android Wear's native notifications requires bezel swiping and scrolling, and we have little control over the text size and layout of a notification. Thus, we designed and implemented a custom watchface to display our prompt messages (\eg, \autoref{fig:design:flow:watchface:prompt}).
We also allowed people to choose either physical or virtual (touchscreen) buttons for most functionalities, considering diverging preferences of older adults on the physical and virtual buttons~\cite{Weisberg2020OlderAdultsWristbands, Manini2019OlderAdultsSmartwatch}. 
We assigned up to two main functions on each screen and placed virtual buttons with a white background (\eg, \autoref{fig:design:flow:watchface:default}--\ref{fig:design:flow:review}) near the top-right and the bottom-right physical buttons on the side, with each virtual button matching the corresponding physical button. 
For consistency, we assigned positive actions (\eg, confirm, launch the reporting) to the top-right button and negative actions (\eg, cancel, dismiss a prompt) to the bottom-right button. 


\phantomsection{}
\label{sec:dr2}
\ipstart{DR2: Simplify Data Capture Flow}
Considering that data entry is repeated frequently, we streamlined the user interface flow for activity reporting. For example, people can submit an entry by pressing the top-right button twice, first to initiate the recording (\autoref{fig:design:flow:watchface:default} or \ref{fig:design:flow:watchface:prompt} $\rightarrow$ \autoref{fig:design:flow:record}), and second to end the recording (\autoref{fig:design:flow:record} $\rightarrow$ \autoref{fig:design:flow:review}). Upon completion of the recording, the review screen (\autoref{fig:design:flow:review}) automatically submits the report so that people do not have to explicitly press the ``OK'' button.
We followed the design of traditional voice recording interfaces, initially allowing pausing and resuming the recording. However, throughout the pilot study we found that pausing/resuming was rarely used but rather made the flow more confusing and therefore removed that functionality.

\phantomsection{}
\label{sec:dr3}
\ipstart{DR3: Leverage the Flexibility of Natural Language Speech Input}
It can be challenging to specify activity types or time/duration information using only graphical user interface widgets on a smartwatch. \revised{The screen is so small that entering data via a text box can be inefficient.} Selecting an activity type from a long list of activities is tedious and prone to error (\eg, ExtraSensory's smartphone app~\cite{Vaizman2018ExtraSensoryApp} supports about 50 activity tags on a hierarchical list, but its companion smartwatch app does not support this tagging activity). Furthermore, specifying time/duration using touch is laborious and inflexible; a smartwatch's small screen does not afford two time pickers (for start and end) in one screen and existing time pickers are not flexible enough to handle the various ways to specify time (\ie, people should specify absolute start and end time)~\cite{Kim2021Data@Hand}.
We also wanted to allow participants to freely describe the effort level to examine what expressions they use to gauge their effort in what situation instead of using the validated scales such as Borg's CR10 scale~\cite{Borg1998BorgScales, Chung2015RPEValidityOlderAdults}.

To mitigate these limitations, 
we leveraged speech input that affords a high level of freedom without requiring much screen space~\cite{Kim2021Data@Hand}. 
People can specify multiple information components in a single verbal report (\eg, ``\textit{I took a 30-minute walk}'' to specify an activity with duration; ``\textit{I did gardening, fixing flower beds from 9:00 to 10:30, in moderate intensity}'' to specify an activity with duration and effort level). 

\subsection{Data Collection}
\label{sec:datacollection}
\ipstart{Verbal Activity Reports}
MyMove collects verbal reports in two different ways: people can submit a report voluntarily at any time or they can submit a report responding to ESM prompts\footnote{Refer to our supplementary video that demonstrates the two reporting methods.}. \revised{Each prompt is scheduled to be delivered at random within hourly time blocks while people are wearing the smartwatch. To send the prompts only when people are wearing the watch,} we leveraged the smartwatch's built-in off-body detect sensor. Once a prompt is delivered, the next one is reserved within the next hour window while leaving at least a 30-minute buffer after the previous one. If the user submits a voluntary report, the next prompt is rescheduled based on the submission time following the same rule. 

We incorporated custom watchfaces to provide coherent visual interfaces (\autoref{fig:design:flow}). 
On the default screen, the watchface displays a clock, the number of reports (logs) that were submitted during the day, and a record button to initiate voluntary reporting (\autoref{fig:design:flow:watchface:default}). When a prompt is delivered, the smartwatch notifies the user with two vibrations and displays a message ``\textit{Describe in detail what you are doing now.}'' with the record and dismiss buttons on the watchface (\autoref{fig:design:flow:watchface:prompt}).
The prompt on the watchface stays for 15 minutes. However, for safety reasons, prompts are skipped if the system recognizes that the user is driving based on the Google Activity Recognition API~\cite{GoogleActivityRecognition}.

When the user starts the recording by tapping on the ``Record'' or \symbolmic button on the watchface (or corresponding physical button), the watch vibrates three times while displaying the message, ``\textit{Start after buzz},'' to indicate initiation. Then MyMove shows the Record screen~(\autoref{fig:design:flow:record}), where people can describe an activity in free-form. The screen displays a message, ``\textit{Activity, duration, and effort level?}'' to remind people of the information components to be included. Recordings can be as long as 2 minutes; after which the session is automatically canceled and the audio is discarded. 
The user completes the recording by pressing the ``End'' button, after which they are sent to the Review screen (\autoref{fig:design:flow:review}) where they can play back the recorded audio (the Fossil watch had a speaker). The recording is submitted upon pressing the ``OK'' button or after 8 seconds without any interaction. 
While recording or reviewing, the user can discard the report using the \symboldelete button.

\ipstart{Background Sensor Data}
MyMove also collects three behavioral and physiological measurements from the onboard sensors and APIs in the background. 
First, every minute, MyMove records a 20-second window of inertial sensor measurements---accelerometer, rotation vector, magnetometer, and gravity---in 25 Hz (500 samples each).
Second, the system records the step counts in one-minute bins and heart rate samples (BPM) at every minute using the smartwatch's built-in sensors. 
Lastly, MyMove collects the classification samples from Google Activity Recognition API, a built-in API that classifies the present locomotion status (\eg, \textit{walking}, \textit{running}, \textit{still in position}, \textit{in vehicle}, \textit{on bicycle}) based on the onboard sensors.

\subsection{Implementation}
We implemented the MyMove app in Kotlin~\cite{Kotlin} on Android Wear OS 2 platform. As a standalone app, it does not require a companion app on the smartphone side.\footnote{The Wear OS 2+ watches can be paired with both iPhone and Android.} 
The verbal reports and sensor data are cached in local storage and uploaded to the server when the smartwatch has a stable internet connection. To optimize network traffic and disk space, the MyMove app serializes sensor data using Protocol Buffers~\cite{Protobuf} and writes them in local files. The server stores the received data in a MySQL database.
\section{Deployment Study}
In May--July 2021, we conducted a deployment study using MyMove to examine the feasibility of speech-based activity labeling on a smartwatch with older adults and the usefulness of the verbal reports in activity labeling. As part of this study, participants reported their activities using a smartwatch while also wearing an activPAL activity monitor~\cite{ActivPAL} on their thigh; this monitor served to collect ground-truth activity data to complement those captured by the wrist worn smartwatch.
Due to the COVID-19 pandemic, all study sessions (introductory, tutorial, and debriefing sessions) were held remotely using Zoom video calls and the study equipment was delivered and picked up by a researcher, complying with COVID-19 prevention guidelines.
This study was approved by the Institutional Review Board of the University of Maryland, College Park. 

\subsection{Pilot Study}
\revised{We iterated on the MyMove design (\eg, data capture flow) and the study procedure (\eg, tutorials) via piloting with two older adults. In an attempt to balance the power structure between older adult participants and our research team, our first pilot participant was a retired HCI researcher. We asked them to follow the study procedure, interact with MyMove and the thigh-worn sensor for 3 days, and provide feedback on the overall study, not as a representative participant but as someone who is both a member of the intended user group and an expert in human-computer interaction. Their feedback informed our design refinement by significantly simplifying the interaction flows, incorporating icons and labels, as well as adding visual feedback making the consequence of users' interactions more noticeable. Upon refining the app design and corresponding tutorial materials, we conducted a second pilot session with another older adult (without any HCI background) to ensure that the watch app and tutorial materials are understandable. }

\subsection{Participants}

We recruited 13 older adults (P1--P13; 10 females and three males) through various local senior community mailing lists in the Northeast region of the United States. Since our study required in-person delivery of the study equipment, we recruited participants in the local area. Our inclusion criteria were adults who (1) are aged 60 or older; (2) feel comfortable describing their activity in English; (3) are curious about their activity levels and interested in collecting activity data; (4) have no severe speech, hearing, motor, movement, or cognitive impairments; (5) have stable home Wi-Fi and are able to join Zoom video calls; and (6) are right-handed. \revised{We exclusively recruited right-handed people because Fossil Gen 5 is designed to be worn on the left wrist. The physical buttons are on the right side of the display with the fixed orientation, making it difficult to maneuver the buttons with the left hand. This also helped to minimize the effect of handedness on sensor data.}

\autoref{table:demographic} shows the demographic information of our study participants \revised{and the average daily activities during the data collection period, measured by activPAL monitors}. All participants were native English speakers and their ages ranged from 61 to 90 (\textit{avg} = 71.08). 
Eight participants were retirees, three were self-employed, and two were full-time employees.
Participants had diverse occupational backgrounds and all participants had Bachelor's or graduate degrees; five had Master's degrees and one had a Ph.D.
All participants were smartphone users; seven used an iPhone and six used an Android phone. 

\revised{
\cameraready{The 7-day activePAL sensor data we collected during the study show our participants' activity level in more detail}: 
Based on existing conventions for interpreting older adults' physical activity volume (i.e., step counts), many of the participants were ``low active'' (46\%; 5000--7499 steps/day) or ``sedentary'' (15\%; $<5000$ steps/day)~\cite{Catrine2011HowManySteps}. The majority of the participants~(77\%) did not meet the 150 min/week of moderate-to-vigorous physical activity (MVPA) recommended in the 2018 Physical Activity Guidelines for Americans~\cite{Piercy2018PhysicalActivityGuidelines}. The average daily physical activity volume ($M$ = 7246.69, $SD$ = 2302.42 steps/day) was consistent with reduced all-cause mortality risk from previous studies with older women~\cite{Lee2019StepVolume}. The mean duration of sedentary behavior was 10 hours and 44 minutes per day ($SD$ = 2 hours and 33 minutes). This high level of sedentary behavior is comparable to device-measured normative values from older adults (10.1 hours/day)~\cite{Rosenberg2020DeviceAssessed} and exceeds self-reported normative values from older adults (6.1 hours/day)~\cite{Yang2019TrendsInSedentary}.
}

\begin{table*}[t]

\caption{Summary of age and gender of our study participants, their employment status and the latest (or current) occupation, education level, technical proficiency, \revised{ and the average daily activities measured with an activPAL monitor during the data collection period, including step count, the time spent for moderate-to-vigorous physical activity (MVPA, the total duration at least 100 steps/min), and the time spent sedentary (the time spent sitting and lying while waking).}}
\vspace{-2mm}
    \small\sffamily
    			\def\arraystretch{1.3}
    		    \setlength{\tabcolsep}{0.4em}
    		    \centering
    \begin{tabular}{|l!{\color{lightgray}\vrule}l!{\color{lightgray}\vrule}ll!{\color{lightgray}\vrule}l!{\color{lightgray}\vrule}l!{\color{lightgray}\vrule}r!{\color{lightgray}\vrule}r!{\color{lightgray}\vrule}l|}
    \hline
    \rowcolor{tableheader}
    \multicolumn{2}{|l!{\color{lightgray}\vrule}}{} & & & & & \multicolumn{3}{l|}{\textbf{\revised{activPAL daily average}}} \\
    \rowcolor{tableheader}
    \multicolumn{2}{|l}{\textbf{Participant}} & \multicolumn{2}{!{\color{lightgray}\vrule}l!{\color{lightgray}\vrule}}{\textbf{Employment \& Latest occupation}} & \textbf{Education} & \textbf{Tech proficiency} & \textbf{\revised{Steps}} & \textbf{\revised{MVPA}} & \textbf{\revised{Sedentary}} \\
    \hline
    \textbf{P1}             & 61 (M)            & Retired & Senior manager                           & Bachelor's & Very confident                 & \revised{10,941} & \revised{<1m} & \revised{11h 23m}     \\
    \arrayrulecolor{tablegrayline}\hline
    \textbf{P2}             & 67 (F)          & Self-employed & Visual artist       & Bachelor's & Enjoy the challenge            & \revised{6,192} & \revised{21m} & \revised{6h 21m}     \\
    \hline
    \textbf{P3}             & 77 (F)          & Retired & Qualitative researcher                            & Ph.D./M.D. & Very confident                 & \revised{9,655} & \revised{2m} & \revised{10h 53m}     \\
    \hline
    \textbf{P4}             & 70 (M)            & Self-employed & Landlord            & Bachelor's & Enjoy the challenge            & \revised{7,793} & \revised{32m} & \revised{9h 7m}         \\
    \hline
    \textbf{P5}             & 81 (F)          & Retired & Disability consultant                            & Master's & A little apprehensive          & \revised{8,773} & \revised{23m} & \revised{7h 48m}         \\
    \hline
    \textbf{P6}             & 79 (F) & Retired & Policy analyst                             & Master's & Very confident                 & \revised{7,320} & \revised{16m} & \revised{9h 12m}          \\
    \hline
    \textbf{P7}             & 69 (F)          & Full-time & Business manager        & Master's & Enjoy the challenge            & \revised{6,499} & \revised{21m} & \revised{12h 5m}         \\
    \hline
    \textbf{P8}             & 90 (F)          & Self-employed & Piano tutor         & Master-level & Enjoy the challenge            & \revised{6,281} & \revised{<1m} & \revised{12h 24m}     \\
    \hline
    \textbf{P9}             & 62 (F)          & Full-time & Communications director  & Master-level  & Very confident                 & \revised{5,313} & \revised{5m} & \revised{13h 50m}     \\
    \hline
    \textbf{P10}            & 62 (F)          & Retired & Human resource specialist                            & Bachelor's & Very confident                 & \revised{3,430} & \revised{<1m} & \revised{13h 37m}     \\
    \hline
    \textbf{P11}            & 67 (F)          & Retired & Technical training manager                            & Master-level & Enjoy the challenge            & \revised{7,296} & \revised{2m} & \revised{7h 19m}         \\
    \hline
    \textbf{P12}            & 75 (F)          & Retired & Rehabilitation counselor                            & Master's & Very apprehensive              & \revised{4,148} & \revised{9m} & \revised{13h 58m}         \\
    \hline
    \textbf{P13}            & 64 (M)            & Retired & Regulatory specialist                            & Master's & Enjoy the challenge            & \revised{10,566} & \revised{46m} & \revised{11h 30m}  \\
    \arrayrulecolor{black}\hline
    \end{tabular}
\label{table:demographic}
\end{table*}

\revised{In appreciation for their participation, we offered participants up to \$150, but we did not tie the activity reporting to the compensation to ensure natural data entry behavior. We provided \$25 for completing the adaptation period with the introductory and tutorial sessions, and another \$25 for a debriefing interview. During the data collection period, we added \$10 for each day of device-wearing compliance (\ie, wear the smartwatch for longer than 4 hours/day), and provided an extra \$30 as a bonus for all seven days of compliance. 
We did not specify a minimum amount of time for wearing the activPAL monitor. 
Compensation was provided after the debriefing session in the form of an Amazon or Target gift card.}

\subsection{Study Instrument}

We deployed a Fossil Gen 5 Android smartwatch, an activPAL4 device, and a Samsung A21 smartphone to each participant. We chose the Fossil Gen 5 Android smartwatch for its large screen size and extended battery life. The smartwatch has a 1.28-inch AMOLED display with a 416 $\times$ 416 resolution (328 PPI). 
To minimize the effort for the initial set up~\cite{Pang2021OlderAdultsTechAdoption}, we deployed smartwatches and Samsung A21 smartphones configured in advance. The phone served as an internet hub for the watch and participants did not have to carry it. While the Bluetooth connection between the watch and the phone was active, the watch periodically uploaded the sensor and verbal reports to our server via the phone's network connection using the participant's home Wi-Fi.

To collect the ground-truth activity postures, we also deployed activPAL4~\cite{ActivPAL}, which is a research-grade activity monitor that uses data from three accelerometers to classify fine-grained body posture and locomotion (\eg, stepping, sitting, lying, standing, in vehicle, and biking). The sensor is attached to the midline of the thigh between the knee and hip using hypoallergenic adhesive tape, and the device does not provide feedback to participants.  
We chose activPAL for three main reasons: First, activPAL can distinguish different stationary postures such as sitting, lying, and standing, more accurately than the wrist-worn or handheld sensors (\eg, Google Activity Recognition API supports only a \textit{Still} class for a stationary state)~\cite{Schrack2016AssessingPASensors}. Second, activPAL is pervasive because it has a long battery life (longer than 3 weeks). 
Third, activPAL yields equivalent reliability to Actigraph devices for physical activity~\cite{Lyden2012activPAL, Kozey-Keadle2011activPAL} and is more accurate than them for capturing slower gait speeds, which are common in older adults~\cite{Ryan2006activPALWalking, Hergenroeder2018StepsOlderAdults}.

\subsection{Study Procedure}

The study protocol consisted of four parts: (1) introductory session and four-day adaptation period, (2) tutorial session, (3) seven-day data collection, and (4) debriefing. We iterated on the study procedure and tutorial materials through the pilot sessions with two older adults.
The introductory, tutorial, and debriefing sessions were held remotely on Zoom. All sessions were recorded using Zoom's recording feature. 


\ipstart{Introductory Session \& Adaptation Period}
After receiving the study equipment, the participant joined a 45-minute introductory session via Zoom. The researcher shared a presentation slide (refer to our supplementary material) via screen sharing. After explaining the goal of the study, the researcher guided the participant to set up the smartphone by connecting it to the home Wi-Fi, wear the smartwatch on the left hand, and attach the activPAL (waterproofed with a nitrile finger cot and medical bandage) to a thigh.
To ensure that the participant felt comfortable handling the smartwatch buttons and the touchscreen elements, we used a custom app in MyMove which can be monitored by the researcher on a web dashboard; the participant went through several trials of pressing a correct button following the message on the screen (\eg, ``\textit{Tap the button [A] on the screen}'' or ``\textit{Push the button [C] on the side}'').

We incorporated the adaptation period to familiarize participants with charging and wearing the devices regularly. During this period, which lasted for four days including the day of introductory session, participants were asked to wear the smartwatch during waking hours and the activPAL for as long as possible. The activity reporting feature was disabled and invisible to the participants. At 9:00 PM, an automated text reminder was sent to participants' own phones to remind them to charge the watch before going to bed.

\ipstart{Tutorial}
On the final day of the adaptation period, we held a 1-hour tutorial session on Zoom to prepare participants for the data collection period starting the next day. The tutorial mainly covered the activity reporting, including a guide on what to describe in a verbal report and how to perform prompted and voluntary reporting with MyMove on a smartwatch. 
We instructed that the verbal reports are ``free-response descriptions about your current or recently-finished activity'' and they can be freely and naturally phrased using one or more sentences. We went through 10 example reports with images of performing the activity in five categories---moving and aerobic exercises, strength exercises, stretching and balance exercises, housekeeping, and stationary activities. All example reports contained the three main information components we are interested in: activity detail, time \& duration, and effort level. For each category, we encouraged participants to come up with imaginary reports including those three components. 

We covered the activity reporting features by demonstrating example flows using animated presentation slides and asking participants to practice on their own watch. Since the session was remote, we observed the participant's smartwatch screen via screen sharing feature of MyMove. We gave participants enough time to practice until they felt comfortable interacting with the smartwatch interface. For the rest of the day, participants were also allowed to submit verbal reports as practice; these reports were not included in the analysis.

\revised{We also explained the compensation rule (see the Participants section above) in detail using a few example cases. We emphasized that the compensation would not be tied to the number of reports, but it would depend on the weartime of the smartwatch (i.e., they need to wear the smartwatch at least 4 hours a day.)} 

\ipstart{Data Collection}
The day following the tutorial, participants started capturing their activities with MyMove, which lasted for one week. During this data collection period, participants received prompt notifications and the device-wearing compliance guideline was in effect. We also sent charging reminders at night just as during the adaptation period.

\ipstart{Debriefing}
After the seventh day of the data collection, we conducted a semi-structured debriefing interview with each participant on Zoom for about 40 to 70 minutes. We asked participants to share their general reactions to the interface and smartwatch as well as their experiences with specifying information components, discussing when they would use prompted or voluntary methods, and if they had a preference towards virtual or physical buttons and why. 
To help participants better recall their experience, we transcribed their verbal reports in advance and shared a summarized table (similar format as \autoref{table:semantics}) via screen sharing.

\revised{Three researchers participated in the debriefing interview sessions, two of whom led the interviews: following the detailed interview script, each researcher covered about a half of the questions. The third researcher observed nine (out of 13) sessions and filled in one session when the second researcher was not available.}



\subsection{Data Analysis}
\label{sec:analysis}
The study produced a rich dataset including the verbal reports that participants submitted, the sensor data captured from the smartwatch and activPAL, and participants' feedback from the debriefing interviews.
We performed both quantitative and qualitative analysis to examine how older adult participants used MyMove to collect in-situ activity labels and to inspect the characteristics and condition of the collected data.
We first examined reporting patterns such as the number of reports collected via two reporting methods as well as audio length and word count of the reports. We analyzed the device usage logs from MyMove and the event logs from activPAL to examine the sensor wearing patterns. 

We then analyzed the transcripts from the verbal reports to understand the semantics of activities participants captured.
Two authors first independently coded a subset of reports after the data collection of the first four participants was completed (80 out of 354; 23\%). We resolved discrepancies and developed the first version of the codebook. As we obtained additional verbal reports from new participants, we iterated multiple sessions of discussions to improve the codebook. After the codebook was finalized, the first author reviewed the entire dataset.
Through a separate analysis, we extracted the effort levels from the reports. Two authors \revised{separately coded a subset of reports (180; 14.7\%) and resolved discrepancies through a series of discussions. After we determined nine categories and how to code data consistently under these categories}, the first author coded the remaining data.  

We further analyzed the transcribed reports to check how diligently participants reported the time component and how well the self-reported information is aligned with the sensor data. We classified the reports into three categories: (1) \textit{No time cues}: the report does not include any time-related information; (2) \textit{Incomplete time cues}: the report includes time cues that are not enough to identify the activity timespan; and (3) \textit{Complete time cues}: the report includes time cues that are sufficient to identify the activity timespan. For example, one of P8's prompted reports, ``\textit{I'm \textbf{just finished} fixing a little dinner.}'' has time-related information (\ie, end time) but we cannot determine the timespan for this activity without the start time or duration. Therefore, this report is classified into the Incomplete time cues category.

\revised{We transcribed the audio recordings of the debriefing interviews. The three researchers who conducted the interviews led the analysis of the debriefing interview data, using NVivo (a qualitative data analysis tool). We grouped the data specific to participants' usability-related experiences with MyMove according to the following aspects: (1) reactions to MyMove and smartwatch, (2) reactions to specifying information components, (3) reactions to using voluntary and prompted methods, and (4) notions on choosing virtual versus physical buttons.}
When appropriate, we also referenced this information while interpreting the results from the analyses mentioned above. 

\section{Results}

\cameraready{We report the results of our study in six parts, aiming to answer the two research questions---first, to demonstrate the feasibility of collecting the activity reports using speech on a smartwatch; and second, to examine the usefulness of the verbal reports as an information source for activity labeling.
In Section \ref{sec:results:descriptive}, we provide an overview of the collected dataset, including participants' engagement in capturing the data. 
In Section \ref{sec:results:activities}, we report the types of activities that participants captured. We specifically discuss how participants' lifestyles and other study contexts affect the reporting patterns and behaviors.
In Section \ref{sec:results:time}, we report how participants describe the time information in their verbal reports and discuss how the nature of an activity and reporting methods affect the completeness of the time cues. We also explore how the verbally-reported activities are aligned with those detected by sensors on a timeline. 
In Section \ref{sec:results:effort}, we explore how participants described their effort level, and assess the validity of the effort level description in relation to the device-based intensity measures.
In Section \ref{sec:results:speechanalysis}, we examine the accuracy of automatic speech recognition technologies in recognizing older adult participants' verbal reports. We further investigate the erroneous instances in detail.
Lastly, in Section \ref{sec:results:experience}, we report on participants' experience with MyMove, based on the qualitative analysis of debriefing interviews.}

\subsection{Dataset Overview}
\label{sec:results:descriptive}

\begin{table*}[t]

\caption{The number of prompted and voluntary reports submitted by each participant. The cell color intensity indicates the ratio between the two reporting methods for each participant.}
\vspace{-2mm}
    \small\sffamily
    			\def\arraystretch{1.2}
    		    \setlength{\tabcolsep}{0.8em}
    		    \centering
\begin{tabular}{cc}
    \begin{tabular}{|l!{\color{lightgray}\vrule}r!{\color{lightgray}\vrule}ccccccccccccc|}
\hline
    \rowcolor{tableheader}
\textbf{Method}  & \textbf{Total} & \textbf{P1} & \textbf{P2} & \textbf{P3} & \textbf{P4} & \textbf{P5} & \textbf{P6} & \textbf{P7} & \textbf{P8} & \textbf{P9} & \textbf{P10} & \textbf{P11} & \textbf{P12} & \textbf{P13} \\

\arrayrulecolor{black}\thicktablehline
\rowcolor{white}
\textbf{Prompted}                                               & 617                                                                 & \cellcolor[HTML]{8DCEBC}32                               & \cellcolor[HTML]{43AE90}66                               & \cellcolor[HTML]{87CCB8}64                               & \cellcolor[HTML]{2AA481}59                               & \cellcolor[HTML]{82CAB5}57                               & \cellcolor[HTML]{D2ECE4}46                               & \cellcolor[HTML]{69BFA6}44                               & \cellcolor[HTML]{AADBCD}33                               & \cellcolor[HTML]{9AD4C3}21                               & \cellcolor[HTML]{2BA482}77                                & \cellcolor[HTML]{ABDBCD}13                                & \cellcolor[HTML]{3CAB8B}55                                & \cellcolor[HTML]{7DC7B2}50                                \\

\textbf{Voluntary}                                              & 607                                                                 & \cellcolor[HTML]{7CC7B1}37                               & \cellcolor[HTML]{C6E7DD}20                               & \cellcolor[HTML]{82C9B5}67                               & \cellcolor[HTML]{DFF1EC}9                                & \cellcolor[HTML]{87CBB8}55                               & \cellcolor[HTML]{37A989}204                              & \cellcolor[HTML]{A0D6C7}28                               & \cellcolor[HTML]{5FBAA0}62                               & \cellcolor[HTML]{6FC1AA}30                               & \cellcolor[HTML]{DEF1EB}12                                & \cellcolor[HTML]{5EBAA0}25                                & \cellcolor[HTML]{CDEAE2}14                                & \cellcolor[HTML]{8CCEBB}44                                \\

\arrayrulecolor{black}\thicktablehline
                 \textbf{Total}  & 1224                   & 69                                            & 86                                           & 131                                          & 68                                            & 112                                          & 250                                            & 72                                           & 95                                           & 51                                           & 89                                            & 38                                           & 69                                            & 94           \\
    \arrayrulecolor{black}\hline
\end{tabular} & 
\begin{tabular}{c}
\includegraphics[height=6.6em]{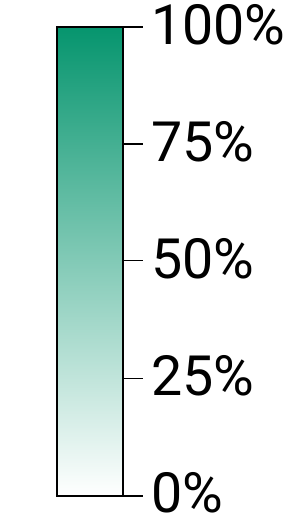}
\end{tabular}
\end{tabular}

\label{table:reports_by_participant}
\end{table*}

While the minimum requirement was to wear the smartwatch and activPAL for at least five days (longer than four hours a day for the smartwatch), all 13 participants wore both devices for the entire seven days. On average, participants wore the smartwatch for 11.6 hours per day ($SD$ = 1.3, $min$ = 9.7~[P11], $max$ = 13.6~[P13]), and activPAL for 23.3 hours per day (10 participants continuously wore activPAL for the entire study period).

We collected 1,224 verbal reports in total, consisting of 617 prompted and 607 voluntary reports: \autoref{table:reports_by_participant} shows the verbal reports by participants. Although the reporting was not tied to the compensation, all participants submitted verbal reports every day. 
Participants submitted 94.15 reports on average, with a high variance among them ($SD$ = 52.85, $min$ = 51~[P9], $max$ = 250~[P6]). 
The average audio length of and word count in each report were 18.65 seconds ($SD$ = 13.65) and 32.05 words ($SD$ = 26.15), respectively. The average audio length per report of each participant ranged from 10.08 [P13] to 32.03 [P12] seconds.

As participants often specified multiple activities in a single report, we extracted \textit{activities} from each report, a unit of continuous task that can be coded with one or (sometimes) two semantics. For example, the report ``\textit{Spent the last 12 minutes, \textbf{eating breakfast}, \textbf{seated in front of the TV}. Minimal level of effort.}'' [P6], specifies two simultaneous activities. 
We identified 1,885 activities from 1,224 verbal reports, and grouped them into the following four categories:
\begin{enumerate}[leftmargin=*]
    \item \textit{Singleton}: 760 (62.10\%) reports contained a single activity,
    \item \textit{Sequential}: 303 (24.75\%) reports contained a series of activities (\textit{avg.} 2.50 activities per report), 
    \item \textit{Multitasking}: 127 (10.38\%) had multiple activities performed simultaneously (\textit{avg.} 2.09 activities per report), 
    \item \textit{Compound}: 34 (2.78\%) were a mix of singleton, sequential, or multitasking (\textit{avg.} 3.06 activities per report). 
\end{enumerate}

\subsection{Captured Activities}
\label{sec:results:activities}

From the 1,885 activities, we identified 29 activity types and grouped them into nine high-level semantics: \textit{housekeeping}, \textit{self-maintenance}, \textit{non-exercise stepping}, \textit{screen time}, \textit{exercise}, \textit{paperwork/desk work}, \textit{hobby/leisure}, \textit{resting}, and \textit{social} (\autoref{table:semantics}). The activity types were generally consistent with prior work in daily activities of older adults~\cite{Horgas1998OlderAdultsLife, Moss1982TimeBudgetOlder}. 
Each participant captured 19.08 unique activity types on average ($SD$ = 4.35, $min$ = 12 [P11], $max$ = 26 [P3]).

\begin{table*}[t]
\caption{Nine activity semantics and 29 activity types, number of reports and participants (Ps), and example snippets from reports. Because the activity semantics and types were multi-coded, the percentages of reports add up to more than 100\%.}
\vspace{-2mm}
\small\sffamily
    			\def\arraystretch{1.1}
    		    \setlength{\tabcolsep}{0.1em}
    		    \centering
\begin{tabular}{!{\color{black}\vrule}p{0.085\textwidth}!{\color{lightgray}\vrule}p{0.13\textwidth}!{\color{gray}\vrule}C{0.03\textwidth}C{0.03\textwidth}!{\color{gray}\vrule}C{0.025\textwidth}!{\color{gray}\vrule}p{0.68\textwidth}!{\color{black}\vrule}}

\arrayrulecolor{black}\hline
\rowcolor{tableheader}\multicolumn{2}{|l}{\textbf{Semantics/types}}                 & \multicolumn{2}{!{\color{gray}\vrule}l!{\color{gray}\vrule}}{\textbf{Reports}} & \textbf{Ps} & \textbf{Example snippet}                                                                                                                                                                         \\

\arrayrulecolor{black}\hline

\multirow[t]{7}{*}{\parbox[t]{2cm}{\vspace{-1.5mm}House-\\keeping}}      & Cleaning/arranging/ carrying  & 263            & 21\%           & 13           & ``\textit{I've been doing some house cleaning which includes vacuuming. And now I'm polishing and dusting.}'' -- P3                                                                                  \\
\arrayrulecolor{lightgray}\cline{2-6}
                                   & Preparing food               & 123            & 10\%           & 13           & ``\textit{I'm in the kitchen and I am just preparing breakfast, so I'm standing at the stove and the toaster.}'' -- P5 \\
\arrayrulecolor{lightgray}\cline{2-6}
                                   & Driving/in a vehicle         & 108            & 9\%            & 12           & ``\textit{Just completed a 30-minute drive, as sitting.}'' -- P1 \\
\arrayrulecolor{lightgray}\cline{2-6}

                                   & Gardening                    & 99             & 8\%            & 11           & ``\textit{I'm picking lettuce in my garden, stooping over. It's not exerting, but it is then a bending and stooping.}'' -- P5 \\
                                   
\arrayrulecolor{lightgray}\cline{2-6}

                                   & Caring for pets              & 68             & 6\%            & 7            & ``\textit{Fed dog, bending over to get food and vegetables and reaching to get pills.}'' -- P6 \\
                                   
\arrayrulecolor{lightgray}\cline{2-6}
                                   
                                   & Offline shopping             & 36             & 3\%            & 11           & ``\textit{At Lowe's, hardware in the garden section. Walking, pushing a stroller and picking up items and plant, for about 40 minutes.}'' -- P3 \\
                                   
\arrayrulecolor{lightgray}\cline{2-6}
                                   
                                   & Other         & 12             & 1\%            & 6            & ``\textit{I have just been doing some light housekeeping chores.}'' -- P7 \\
                                   
\arrayrulecolor{black}\hline
                                   
\multirow[t]{4}{*}{\parbox[t]{2cm}{\vspace{-1.5mm}Self-\\maintenance}}  & Eating food                  & 186            & 15\%           & 13           & ``\textit{Ate breakfast from 6:30 until 7:03.}'' -- P13  \\

\arrayrulecolor{lightgray}\cline{2-6}

                                   & Dressing                     & 36             & 3\%            & 9            & ``\textit{Process of getting dressed for the day. Pulling my clothes together and getting ready for what I'm going to do today.}'' -- P10 \\
\arrayrulecolor{lightgray}\cline{2-6}
                                   & Personal hygiene             & 24             & 2\%            & 8            & ``\textit{Just completed a shower.}'' -- P6 \\
                                   
\arrayrulecolor{lightgray}\cline{2-6} 
                                   
                                   & Treatment                    & 10             & 1\%            & 6            & ``\textit{From 11:45 to 12:45 I had a massage. So I was laying down and there was no intensity level whatsoever.}'' -- P9 \\
                            
\arrayrulecolor{black}\hline

\multicolumn{2}{|l|}{Non-exercise stepping}                         & 171            & 14\%           & 12           & ``\textit{I'm just walking up the stairs to just do some minor things.}'' -- P5 \\

\arrayrulecolor{black}\hline

\multirow[t]{4}{*}{Screen time}       & Computer        & 164            & 13\%           & 11           & ``\textit{I'm on the computer. I'm looking at all the sales offers.}'' -- P12 \\

\arrayrulecolor{lightgray}\cline{2-6}

                                   & TV              & 151            & 12\%           & 12           & ``\textit{I'm watching TV, just I've been watching it for maybe 10 minutes so far.}'' -- P2 \\
                                   
\arrayrulecolor{lightgray}\cline{2-6}
                                   
                                   & Mobile device   & 27             & 2\%            & 4            & ``\textit{I'm sitting, looking at a webinar on the phone.}'' -- P4 \\
                                   
\arrayrulecolor{lightgray}\cline{2-6}
                                   
                                   & Device unspecified     & 17             & 1\%            & 5            & ``\textit{I am sitting in watching videos on YouTube.}'' -- P10 \\

\arrayrulecolor{black}\hline

\multirow[t]{4}{*}{Exercise} & Cardio              & 118             & 10\%            & 11           & ``\textit{I just returned from a 30 minute walk, fairly easy paced, moderate effort because of the heat and humidity.}'' -- P7 \\

\arrayrulecolor{lightgray}\cline{2-6}

                                   & Strength/stretching             & 51             & 4\%            & 8            & ``\textit{I am doing stretching exercises in preparation for my strength training class, which I will be taking and I've been doing stretching for about 10 minutes.}'' -- P10 \\
                                   
\arrayrulecolor{lightgray}\cline{2-6}

                                   
                                   
                                   & Other                       & 10             & 1\%            & 4            & \parbox[t]{10cm}{``\textit{I've just finished an hour and a half long workshop on meditation.}'' -- P3\\ ``\textit{In the 4th hole in the golf course, do playing golf.}'' -- P1 \vspace{0.5mm}} \\
                                   
\arrayrulecolor{black}\hline

\multicolumn{2}{|l|}{Paperwork/desk work}                           & 68             & 6\%            & 10           & ``\textit{balancing my checkbook and writing checks for bills.}'' -- P12                                                                                                      \\
\arrayrulecolor{black}\hline

\multirow[t]{5}{*}{\parbox[t]{2cm}{\vspace{-1.5mm}Hobby/\\leisure}}     & Reading on paper             & 59             & 5\%            & 10           & ``\textit{I'm lying on my bed, reading a book. I've been doing that for half an hour.}'' -- P2  \\

\arrayrulecolor{lightgray}\cline{2-6}

                                   & \parbox[t]{2cm}{Playing puzzle/\\table game}    & 17             & 1\%            & 6            & ``\textit{I'm sitting at the counter in the kitchen doing a Sudoku.}'' -- P5 \\
                                   
\arrayrulecolor{lightgray}\cline{2-6}
                                   
                                   & Crafting/artwork             & 15             & 1\%            & 4            & ``\textit{I've been working, doing some woodworking in the basement. -- P13}'' \\
                                   
\arrayrulecolor{lightgray}\cline{2-6}
                                   
                                   & Seeing at a theater          & 11             & 1\%            & 3            & ``\textit{I've been seated at a concert for the past two hours.} -- P5 \\
                                   
\arrayrulecolor{lightgray}\cline{2-6}
                                   
                                   & \parbox[t]{2.5cm}{Playing a musical\\instrument\vspace{1mm}} & 8              & 1\%            & 2            & ``\textit{I am sitting at my piano, playing the piano.} -- P10  \\
                                   
\arrayrulecolor{black}\hline

\multirow[t]{2}{*}{Resting}         & Nothing/waiting              & 54             & 4\%            & 12           & ``\textit{For the last two hours, I've been sitting, getting my car serviced.} -- P9 \\

\arrayrulecolor{lightgray}\cline{2-6}

                                   & Napping                      & 19             & 2\%            & 7            & ``\textit{Since the last ping I took about half an hour nap.}'' -- P7 \\

\arrayrulecolor{black}\hline

\multirow[t]{2}{*}{Social}            & \parbox[t]{2.5cm}{Face-to-face\\interaction\vspace{1mm}}     & 39             & 3\%            & 9            & ``\textit{I just sat down on my front porch swing and I'm talking to a friend.}'' -- P3 \\

\arrayrulecolor{lightgray}\cline{2-6}

                                   & Voice call                   & 36             & 3\%            & 8            & ``\textit{I just completed a telephone call, regarding a personal business.}'' -- P6 \\

\arrayrulecolor{black}\hline

\end{tabular}

\label{table:semantics}
\end{table*}

Participants frequently captured housekeeping activities such as \textbf{cleaning}, \textbf{arranging} or \textbf{carrying} items. These activities included straightening rooms, vacuuming, washing the dishes, or carrying goods purchased from shopping.
Twelve out of 13 participants were living in a house with a yard and 11 of them captured \textbf{gardening} activities. However, specific tasks varied, ranging from light activities (\eg, watering flowers) to heavy activities (\eg, fixing flower beds, planting trees).
Participants also frequently captured \textbf{non-exercise stepping}, which involves a lightweight physical activity, mostly brief in nature. For example, these activities included going up \& down the stairs, walking around the kitchen at home, and walking to/from a car, as well as pushing a shopping cart in a store. 
Eleven participants regularly engaged in \textbf{cardio exercise}, which includes walking, biking, and swimming. The most common exercise was taking a walk (including walking the dog) whereas more strenuous exercise such as running was rarely captured. Eight participants engaged in \textbf{strength and stretching exercises}, for example, online yoga classes. 
Participants also captured brief strength and stretching exercises (\eg, leg lifts) they performed during other stationary activities such as TV watching or artwork. 
Other types of exercises included online meditation sessions, breathing exercises, and golf. 

During debriefing, participants mentioned factors that affected their engagement in specific activities. 
Gardening was often affected by the season and weather. For example, P1, who participated in the study in mid May, noted that he engaged in gardening more than usual: ``\textit{This was a high active seven days for me [sic]. Both because of weather and the time of year, we’re trying to transition the garden.}'' In contrast, P9, who participated in the study in late June, seldom captured gardening and noted, ``\textit{It was really hot, stinky hot and, you know, not a fun thing to do [gardening] (...) in the earlier in the spring when I planted all my flowers and stuff, that feels more like gardening.}'' 
In addition, the COVID-19 lockdown reduced the overall engagement in outdoor physical activities and in-person activities. P4 noted, ``\textit{I would bike downtown two or three times a week anyhow. Normally if before COVID, I've been down maybe four or five times for the last year.}'' Similarly, P11 remarked, ``\textit{In pre-COVID, I would have done that [swimming] probably twice, two or three times during the week.}'' 
Many participants were involved in one or more community activities and their meetings transitioned to Zoom due to the lockdown, possibly increasing their screen time in place of the face-to-face interactions.

We learned that some activities were inherently easier to capture than others due to the contexts in which they are performed: this may have led to oversampling of those activities. 
For example, P3 commented on her high number of reports of watching TV: ``\textit{That [watching TV] had so many times because I was sitting down and it was easy to use the watch. You know, I was taking a break, and the break allowed me to do that.}''
In addition, common activities were likely to be overlooked, thereby affecting the data capture behavior.
For example, P11, who lives with her grandchildren, noted that she did not capture face-to-face interactions with them because such events happened throughout the day, which makes it overwhelming to capture all of them thoroughly: ``\textit{If I recorded what I do with my grandkids, I would be recording all day [laughs]. A lot of times that I interact with my grandkids is kind of in short verse.}'' 


\subsection{Reporting Patterns for Time}
\label{sec:results:time}

\autoref{table:activities_by_time_cues} summarizes the time cue categories of activities from \textit{Singleton}, \textit{Sequential}, and \textit{Multitasking} reports. We excluded 34 \textit{Compound} reports (104 activities) because it was infeasible to reliably extract time cues for each activity. Overall, 984 out of 1781 activities (55.25\%) were mapped with time cues, and 770 of them (78.25\%) were mapped with \textit{Complete time cues}. The remaining 796 activities (44.69\%) were not mapped with any time cues.

\cameraready{Reports containing a single activity were more likely to include Complete time cues than reports containing multiple activities}: 64.87\% (493/760) of \textit{Singleton} activities were mapped with Complete time cues, compared with 20.11\% (152/756) for \textit{Sequential} and 47.17\% (125/265) for \textit{Multitasking}. Of the 319 activities from \textit{Sequential} activities with time cues, about a half (167) were mapped with \textit{Incomplete time cues} because participants often specified the start and end time of the entire sequence 
(\ie, the start time of the first activity and the end time of the last activity). \cameraready{However, this pattern was not consistent across all participants, mainly due to the high individual variance in the number of total reports (See \autoref{table:reports_by_participant}) and in the portions of \textit{Singleton}, \textit{Multitasking}, and \textit{Compound} activities.} 

\begin{table*}[t]

\caption{Number of activities in \textit{Singleton}, \textit{Sequential}, and \textit{Multitasking} reports by reporting method and the time cue category.}
\vspace{-2mm}

\small\sffamily
    			\def\arraystretch{1.3}
    		    \setlength{\tabcolsep}{0.6em}
    		    \centering
\begin{tabular}{|c|cc!{\color{lightgray}\vrule}c!{\vrule width 0.5pt}cc!{\color{lightgray}\vrule}c!{\vrule width 0.5pt}cc!{\color{lightgray}\vrule}c|}
\hline
\rowcolor{tableheader}          & \multicolumn{3}{c!{\vrule width 0.5pt}}{\cellcolor[HTML]{dedede}\textbf{\textit{Singleton} reports (=activities)}}         & \multicolumn{3}{c!{\vrule width 0.5pt}}{\cellcolor[HTML]{dedede}\textbf{\textit{Sequential} activities}}       & \multicolumn{3}{c|}{\cellcolor[HTML]{dedede}\textbf{\textit{Multitasking} activities}}         \\
\arrayrulecolor{black}\cline{2-10}
\rowcolor{tableheader}          & \multicolumn{2}{c!{\color{lightgray}\vrule}}{With time cue} &           & \multicolumn{2}{c!{\color{lightgray}\vrule}}{With time cue} &            & \multicolumn{2}{c!{\color{lightgray}\vrule}}{With time cue} &          \\
\arrayrulecolor{darkgray}\cline{2-3}\cline{5-6}\cline{8-9}
\rowcolor{tableheader} \textbf{Method} & Complete       & Incomplete       & No cues   & Complete       & Incomplete       & No cues    & Complete       & Incomplete       & No cues  \\
 
\arrayrulecolor{black}\thicktablehline
 
\textbf{Prompted}  & 226            & 14               & 131         & 67             & 59               & 211  & 68             & 4                & 104     \\
\arrayrulecolor{lightgray}\hline
\textbf{Voluntary} & 267            & 27               & 95         & 85             & 108              & 226   & 57             & 3                & 29   \\

\arrayrulecolor{black}\thicktablehline

\textbf{Total}     & 493            & 41               & 226        & 152            & 167              & 437  & 125             & 7                & 133  \\

\hline
\end{tabular}
\label{table:activities_by_time_cues}
\end{table*} 

\cameraready{Voluntary reports were more likely to include Complete time cues than prompted reports}: 45.60\% (409/897) of activities from voluntary reports were mapped with Complete time cues, whereas 40.84\% (361/884) from prompted reports. Participants were more likely to omit time cues in prompted reports, especially when reporting simultaneous activities: 61.36\% (108/176) of \textit{Multitasking} activities from prompted reports contained Incomplete or No time cues, in comparison with 35.94\% (32/89) of those from voluntary reports. \cameraready{Again, these patterns were not consistent across participants with high individual variance.} 

\begin{figure*}[b]
    \centering
    \includegraphics[width=\textwidth]{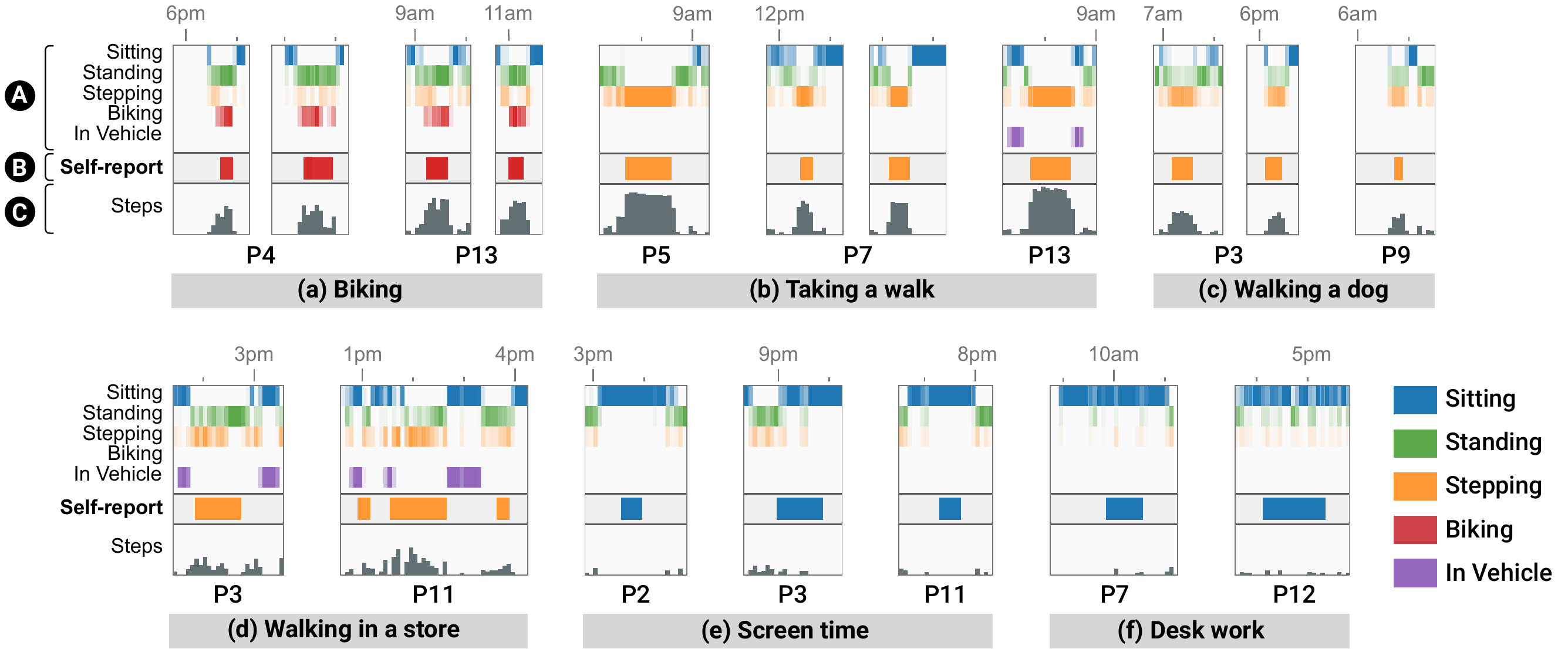}
    \labelphantom{fig:timeline:biking}
    \labelphantom{fig:timeline:walk}
    \labelphantom{fig:timeline:dog}
    \labelphantom{fig:timeline:store}
    \labelphantom{fig:timeline:screen}
    \labelphantom{fig:timeline:paperwork}
    \vspace{-3mm}
    \caption{The excerpts of self-report time segments (\circledigit{B}) of selected activities, along the timeline with automatically-inferred activities (\circledigit{A}) and step counts (\circledigit{C}) from activPAL. The colors denote the types of activPAL's activity classes. The self-report time segments are color-coded as the equivalent activPAL classes.
    }
    \label{fig:timeline}
    \Description{
        A set of timeline visualizations, juxtaposing the activity timeline of activPAL, self-report time segments, and the distribution of step counts from activPAL. There are six subfigures for each selected activity. (a) Biking shows a subset of activities reported by P4 and P13 (referred to as red segments in the text). (b) Taking a walk shows a subset of activities reported by P5, P7, and P13 (referred to as orange segments in the text). (c) Walking a dog shows a subset of activities reported by P3 and P11 (referred to as orange segments in the text). (d) Walking in a store shows a subset of activities reported by P3 and P11 (referred to as orange segments in the text). (e) Screen time shows a subset of activities reported by P2, P3, and P11 (referred to as blue segments in the text). (f) Desk work shows a subset of activities reported by P7, and P12 (referred to as blue segments in the text).
    }
\end{figure*}

Regarding the reports with \textit{Complete time cues}, we investigated how time segments from verbal reports are aligned with those detected by activPAL. 
\autoref{fig:timeline} shows the excerpts of timelines with self-report time segments of selected activities, along with the inferred activities and step counts from activPAL.
Time segments from verbal reports for locomotion-based cardio exercises such as biking and taking a walk generally corresponded with the bands with an equivalent activPAL class and clusters of peaks in step counts. \revised{For example,} the red segments in \autoref{fig:timeline:biking} and orange segments in \autoref{fig:timeline:walk} \revised{illustrate how they are} aligned with activPAL's \textit{Biking} and \textit{Stepping} bands. 
Other kinds of walking activities from verbal reports, such as walking a dog and moving in a store also corresponded with the activPAL activity patterns, but participants' movement was more fragmented with the \textit{Standing} and \textit{Stepping} classes compared to a pure walking exercise (see the orange segments in \autoref{fig:timeline:dog} and \ref{fig:timeline:store} which also overlap with activPAL's \textit{Standing} band).

Activities performed while sitting often did not correspond with the momentary changes in the activPAL activities. For example, blue segments in \autoref{fig:timeline:screen} and \ref{fig:timeline:paperwork} indicate screen time and desk work activities that participants reported performing while sitting. In all cases, the bands of activPAL's \textit{Sitting} class cover a wider region than the self-report time segments.

\subsection{Reporting Patterns for Effort Level}
\label{sec:results:effort}

About a half of reports (644 out of 1,224 reports, 52.61\%) contained cues on the effort level (see \autoref{table:reports_by_effort}), with high variance among participants ($SD$ = 31.19\%; $min$ = 5.26\%~[P8], $max$ = 98.04\%~[P9]).
We grouped the effort level cues into seven orderly categories on a spectrum of \textit{No effort}--\textit{Low}--\textit{Moderate}--\textit{Strenuous}, and two additional categories---\textit{Relaxed} and \textit{Uncategorizable} (see \autoref{table:effort:categories}). 
The most common effort level reported were \textit{Low} activities (276 reports by 12 participants), followed by \textit{Moderate} activities (132 reports by 11 participants). The majority of \textit{Low} activities were stationary activities such as screen time, eating, driving, or desk work, and the \textit{Moderate} activities included exercises, gardening, or thorough cleaning activities. \textit{Strenuous} activities were rarely captured (20 reports by 5 participants).  
The \textit{Relaxed} category includes responses such as ``\textit{I'm sitting \textbf{totally relaxed}, reading my phone and watching TV}'', and the \textit{Uncategorizable} category covers responses that conveyed ambiguous level of effort (\eg, ``\textit{Stretches for my back, knee bends. \textbf{Nothing too strenuous} but just to break up the sitting.}'').

\begin{table*}[t]
\caption{Number of reports with and without effort level cues by each participant.}
\vspace{-2mm}
    \small\sffamily
    			\def\arraystretch{1.2}
    		    \setlength{\tabcolsep}{0.8em}
    		    \centering
\begin{tabular}{cc}
\begin{tabular}{|l!{\color{lightgray}\vrule}r!{\color{lightgray}\vrule}ccccccccccccc| }

\arrayrulecolor{black}\hline

\rowcolor{tableheader}
\textbf{Effort level cue}              & \textbf{Total} & \textbf{P1}                & \textbf{P2}                & \textbf{P3}                 & \textbf{P4}                & \textbf{P5}                & \textbf{P6}                 & \textbf{P7}                & \textbf{P8}                & \textbf{P9}                & \textbf{P10}               & \textbf{P11}               & \textbf{P12}               & \textbf{P13}               \\

\arrayrulecolor{black}\thicktablehline
\rowcolor{white}

\textbf{Included}                                & 644            & \cellcolor[HTML]{A6D8CA}25 & \cellcolor[HTML]{29A380}75 & \cellcolor[HTML]{D9EEE8}20 & \cellcolor[HTML]{8FCFBC}30 & \cellcolor[HTML]{BDE2D8}30 & \cellcolor[HTML]{53B599}175 & \cellcolor[HTML]{70C1AA}42 & \cellcolor[HTML]{F2F9F7}5  & \cellcolor[HTML]{0E9870}50 & \cellcolor[HTML]{179B76}84 & \cellcolor[HTML]{50B497}27 & \cellcolor[HTML]{3BAB8B}55 & \cellcolor[HTML]{BDE3D8}25 \\
\textbf{Not included}                               & 580            & \cellcolor[HTML]{62BCA2}44 & \cellcolor[HTML]{DFF1EC}11 & \cellcolor[HTML]{2FA684}111 & \cellcolor[HTML]{79C5B0}38 & \cellcolor[HTML]{4BB294}82 & \cellcolor[HTML]{B5DFD3}75  & \cellcolor[HTML]{98D3C2}30 & \cellcolor[HTML]{169B75}90 & \cellcolor[HTML]{FAFCFC}1  & \cellcolor[HTML]{F1F9F6}5  & \cellcolor[HTML]{B8E0D5}11 & \cellcolor[HTML]{CDE9E1}14 & \cellcolor[HTML]{4BB194}69 \\

\arrayrulecolor{black}\thicktablehline

\textbf{Total} & 1224           & 69                         & 86                         & 131                         & 68                         & 112                        & 250                         & 72                         & 95                         & 51                         & 89                         & 38                         & 69                         & 94 \\

\arrayrulecolor{black}\hline

\end{tabular} &

\begin{tabular}{c}
\includegraphics[height=6.6em]{figures/table-color-legend-vert.pdf}
\end{tabular}

\end{tabular}

\label{table:reports_by_effort}
\end{table*}

\begin{table*}[b]

\caption{Categories of verbalized effort level cues with the number of reports and participants (Ps), and example phrasings from utterances. The effort level cues are highlighted in bold.}
\vspace{-2mm}
    \small\sffamily
    			\def\arraystretch{1.4}
    		    \setlength{\tabcolsep}{0.2em}
    		    \centering
\begin{tabular}{!{\color{black}\vrule}p{0.17\textwidth}!{\color{lightgray}\vrule}C{0.06\textwidth}!{\color{lightgray}\vrule}C{0.025\textwidth}!{\color{lightgray}\vrule}p{0.71\textwidth}!{\color{black}\vrule}}

\hline
    \rowcolor{tableheader}

\textbf{Effort level category} & \textbf{Reports} & \textbf{Ps} & \textbf{Example phrasings}                   \\

\hline

\textbf{Relaxed} & 43                & 9                      & \parbox[t]{0.64\textwidth}{``\textit{Lying in bed, watching a retirement seminar life. \textbf{Super relaxed}.}'' -- P4 \vspace{1.5mm}\\ 
``\textit{I'm sitting down and the salesperson is helping me try on shoes. \textbf{Pretty leisurely}.}'' -- P3} \\
\arrayrulecolor{tablegrayline}\hline
\textbf{No   effort}              & 87                & 8                      & ``\textit{Trying to research something on my computer. \textbf{No effort}.}'' -- P2                   \\
\hline
\textbf{No-to-Low}            & 5                 & 3                      &  ``\textit{Standing in the kitchen, preparing lunch. \textbf{Little to no effort}.}'' -- P1                             \\
\hline
\textbf{Low}        & 276               & 12                     & \parbox[t]{0.64\textwidth}{``\textit{I've been eating for probably about 20 minutes. And \textbf{effort level is low}.}'' -- P10 \vspace{1mm}\\
``\textit{Had a 15 minute walk with the dog. It was \textbf{light exertion}.}'' -- P9 \vspace{1mm}\\
``\textit{I've been in the kitchen, cooking. \textbf{Minimal effort}.}'' -- P7} \vspace{1mm} \\
\hline
\textbf{Low-to-Moderate}        & 37                & 5                      & ``\textit{Cutting material for large raised bed garden. \textbf{Light to moderate activity}.}'' -- P6                              \\
\hline
\textbf{Moderate}  & 132               & 11                     & \parbox[t]{0.64\textwidth}{``\textit{In the garden again and bending down, digging holes in the ground. \textbf{Moderate exertion}.}'' -- P2 \vspace{1mm}\\
``\textit{Thoroughly wiped down stainless refrigerator and cleaned inner seal of doors, 25 minutes.\\ \textbf{Medium exertion}.}'' -- P6 \vspace{1mm} \\
``\textit{Preparing lunch, heating a bowl of soup up. \textbf{My activity level is average}.}'' -- P10}\vspace{1mm} \\
\hline
\textbf{Moderate-to-Strenuous}  & 10                & 2                      & ``\textit{Walking through the airport for about a half hour, \textbf{medium to heavy intensity}.}'' -- P9                              \\
\hline
\textbf{Strenuous}                & 20                & 5                      & \parbox[t]{0.64\textwidth}{``\textit{I moved boxes and canned goods and so on into the storage area. \textbf{Expended a great}\\ \textbf{deal of energy doing that. Was tired afterwards}.}'' -- P12}\vspace{1mm}  \\
\hline
\textbf{Uncategorizable}                    & 44                & 8  & \parbox[t]{0.64\textwidth}{``\textit{Dressing and cleaning up for about 15 minutes total. \textbf{Not much effort}.}'' -- P5} \\      

\arrayrulecolor{black}\hline
\end{tabular}
\label{table:effort:categories}
\end{table*}

To examine how self-report effort level categories are related with device-based intensity measures, we compared intensity measurements across the effort level categories using \textit{mixed-effects models} because these models can handle unbalanced data with repeated measured from the same participant~\cite{Pinheiro2000MixedEffects}.
For this analysis, we included 480 activities that contained both Complete time cues and Effort level cues; we counted two or more activities included in \textit{Multitasking} reports as one activity because multiple activities (\eg, ``\textit{watching TV while eating dinner}'') were mapped to one effort level (\eg, ``\textit{it was very low effort}''). In this analysis, we excluded the \textit{Uncateogrizable} category.
We employed two common indicators of intensity in physical activity research---\textit{the percentage of HR$_{max}$} (the average heart rate during the period expressed as a percentage of age-adjusted maximum heart rate\footnote{We used Nes and colleagues' formula ($211-0.64*age$)~\cite{Nes2013HeartRate} as an estimate of age-adjusted maximum heart rate to reflect the age-related changes.}) and \textit{walking cadence} (steps/min)~\cite{Tudor-Locke2012Cadence, acsm2002exercise}. We generated a model for each of the three measurements---the percentage of HR$_{max}$ from smartwatch, walking cadence from activPAL, and walking cadence from smartwatch.
We used intercept (participant) as a random effect and effort level category as a fixed effect. From Maximum-likelihood tests with other variables, we found that \textit{age}, \textit{elapsed days}, and \textit{activity types} did not have significant effects on the measurements. Therefore, we excluded them from fixed effects in the models.

\begin{figure*}[t]
    \begin{flushleft}
    \sffamily\footnotesize{\textbf{***}$p$ < .001; \textbf{**}$p$ < .01; \textbf{*}$p$ <. 05}
    \end{flushleft}

    \centering
    \includegraphics[width=\textwidth]{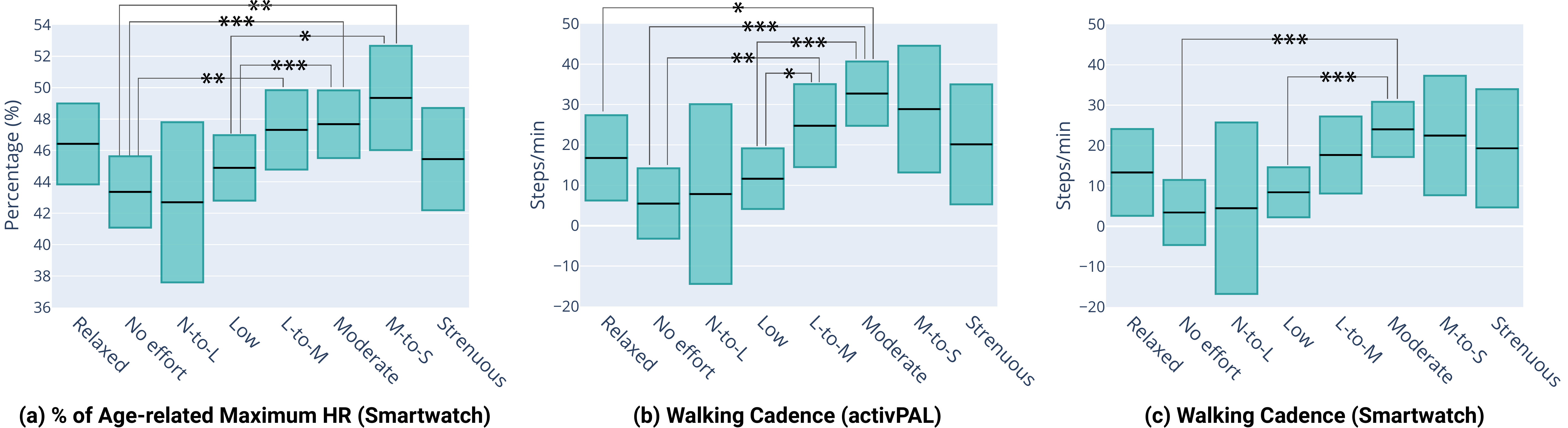}
    
    \labelphantom{fig:chart:effort:metrics:hr}
    \labelphantom{fig:chart:effort:metrics:stepActivPal}
    \labelphantom{fig:chart:effort:metrics:stepWatch}
    \vspace{-5mm}
    
    \caption{Distributions of device-based intensity measurements during the same time segments for each effort-level category. The colored rectangles denote 95\% confidence intervals estimated by the mixed-effects model with a center bar as the least squared mean after controlling the individual differences. The asterisks with arms indicate significance between the connected categories. (We did not mark the pairs that are not significant.)}
    \label{fig:chart:effort:metrics}
    \Description{
        Containing three subfigures for (a) the percentage of HR$_{max}$, (b) walking cadence from activPAL, and (c) walking cadence from smartwatch. Each subfigure shows the sensor-based measurements on the Y axis against self-report effort level categories on the X axis. There is an increasing trend in the measurements from No effort to Moderate-to-Strenuous activities. In all cases, the measurements of Strenuous activities are plotted lower than Moderate or Moderate-to-Strenuous activities.
    }
\end{figure*}

We found significant differences among the effort level categories in their intensity measurements across all three metrics: $F$(7, 407.69) = 7.32, $p$ < .001 for the percentage of HR$_{max}$; $F$(7, 446.69) = 12.00, $p$ < .001 for walking cadence from activPAL; and $F$(7, 369.96) = 6.19, $p$ < .001 for walking cadence from the smartwatch.
We conducted post-hoc pairwise comparisons of the least-squared means of intensity measurements among 8 effort level categories using Tukey adjustment in \texttt{emmeans}~\cite{emmeans} package in R. 
\autoref{fig:chart:effort:metrics} visualizes the significance over the 95\% confidence intervals of measurements in each category. 
Across all three metrics, the intensity measurements of the activities specified as \textit{Moderate} were significantly higher than those of \textit{No effort} ($p$ < .001) and \textit{Low} ($p$ < .001). 
The percentage of HR$_{max}$ and activPAL-measured walking cadence for \textit{Low-to-Moderate} activities were also significantly higher than those of \textit{No effort} activities ($p$ = .005 for the percentage of HR$_{max}$ and $p$ = .004 for walking cadence).
For \textit{Moderate-to-Strenuous} activities, only the percentage of HR$_{max}$ was significantly higher than that of \textit{No effort} ($p$ = .003) and \textit{Low} ($p$ = .036) activities. The activities specified as \textit{No effort} and \textit{Low} did not differ across all metrics. 

Participants' subjective evaluation of the effort level did not match the standard intensity level of physical activity, especially for the activities that are Moderate or above (26.67\%; 128/480).   
Of the 119 \textit{Moderate}, \textit{Moderate-to-Strenuous}, and \textit{Strenuous} activities with the percentage of HR$_{max}$ measurements, only one activity exceeded the lower bound of standard moderate intensity (64\%--76\% for moderate-intensity physical activity~\cite{acsm2002exercise}). Similarly, five (out of 128) and three (out of 113) activities in the same categories exceeded the threshold of moderate intensity walking cadence (100 steps/min or higher for moderate activity~\cite{Tudor-Locke2012Cadence}) with the measurements from activPAL and the smartwatch, respectively.

\begin{table}[b]
\caption{\revised{Regression model for the effort level score, fitted from the device-based intensity measurements, $F$(3, 345) = 15.25, $p$ < .0001, adjusted $R^2$ = .11. The positive coefficient denotes that the given parameter is positively correlated to the effort level score.}}
\vspace{-2mm}
    \small\sffamily
    			\def\arraystretch{1.4}
    		    \setlength{\tabcolsep}{0.5em}
    		    \centering
\begin{tabular}{|l!{\color{lightgray}\vrule}llll|}

\hline
    \rowcolor{tableheader}
                                    \textbf{Parameter} & \textbf{Coef.}  & $\bm{SE}$   &  $\bm{t}$\textbf{-statistic}     & {$\bm{p}$\textbf{-value}}                   \\
\hline
Constant                             & \textcolor{white}{-}2.00  & 0.75 & \textcolor{white}{-}2.67  & \textless{} .01**   \\
\arrayrulecolor{tablegrayline}\hline
Walking cadence (activPAL)       & \textcolor{white}{-}0.02  & 0.01 & \textcolor{white}{-}3.55  & \textless{} .001*** \\
\hline
Walking cadence (Smartwatch)       & -0.01 & 0.01 & -1.43 & .15                \\
\hline
\% of age-related maximum HR & \textcolor{white}{-}0.03  & 0.02 & \textcolor{white}{-}1.71  & .09      \\
\arrayrulecolor{black}\hline

\end{tabular}
    \begin{flushleft}
    \hspace{53mm}
    \sffamily\footnotesize{
    \textbf{***}$p$ < .001; \textbf{**}$p$ < .01; \textbf{*}$p$ < .05}
    \end{flushleft}
    
\label{table:effort:regression}
\end{table}

\revised{To examine how predictive the device-based intensity measurements are for the effort level, we conducted a multiple linear regression analysis using \texttt{MASS}~\cite{mass} package in R. This method initially adds all predictors---the three device-based intensity measurements---to a model and iteratively excludes the predictors that do not make a significant contribution to the prediction, reassessing the contributions of the remaining predictors at each step. We first transformed the seven ordinal categories (\textit{No effort}--\textit{Strenuous}) into a continuous effort level scale (1--7, with \textit{Low} as 3 and \textit{Moderate} as 5) and used it as a dependent variable. For this analysis, we included 349 activities which contain the values of all three measurements.
A significant regression equation (see \autoref{table:effort:regression}) was found ($F$(3, 345) = 15.25, $p$ < .0001), with an adjusted $R^2$ of .11. Although all three measurements collectively contributed to the prediction and were thus included to the final model, only walking cadence from activPAL was statistically significant~($p$ = .0004). The $R^2$ value denotes that the model explains only 11\% of the variance of the effort level scores. This implies that it may not be feasible to accurately predict the exact effort level score using only the device-based measurements.
}

\subsection{Quality of Voice Recording}
\label{sec:results:speechanalysis}
To investigate the potentials of activity labeling with speech input, we assessed how accurately the existing automatic speech recognition (ASR) technologies can recognize participants’ speech inputs, especially since there is prior evidence on disproportionate ASR word error rates for older adults' voices~\cite{Chen2021ASROlderAdults, vipperla2008longitudinal}.  
Considering the transcribed text of verbal reports by our research team as the ground-truth, we compared it with the output from two commercial ASR services, Microsoft Cognitive Speech~\cite{MicrosoftCognitiveSpeech} and Google Cloud Speech~\cite{GoogleCloudSpeech}. Using their REST APIs, we retrieved the recognized text from the audio files for each verbal report. We then calculated Word Error Rate (WER) of the recognized text using the human-transcribed text. When calculating WER, we removed punctuation and fixed contractions using NLTK (Natural Language Toolkit)~\cite{nltkbook} and Contractions Python Library~\cite{ContractionsLib}. 
On average, the Microsoft API recognized reports with an word error rate of 4.93\% per report per participant ($N$ = 13, $SD$ = 2.12\%). This is slightly lower than 5.10\% that Microsoft had reported in 2018~\cite{Xiong2018MicrosoftAsr}. The Google API yielded an error rate of 8.50\% per report per participant ($N$ = 13, $SD$ = 2.97\%). This is 3.60\% higher than 4.90\% that Google had officially announced in 2017~\cite{Protalinski2017GoogleAsr}. 

\revised{We performed an error analysis to gain insights into the potential effect these errors may have in automating the retrieval of activity labels from free form verbal reports. Specifically, we manually inspected a total of 651 verbal reports where there was a disagreement between our ground truth and the best performing ASR service. Many of the errors (70.97\%; 462/651) did not affect the words capturing activity type, time, or effort level, \ie, with the local context of the verbal report someone could correctly infer this information if it was reported. Typically, errors in these reports involved filler words, conjunctions, or other details that participants provided along their activity. 
For example, misrecognized conjunction in the ASR output of P1's report, ``\textit{Eating lunch, \textbf{Ann} [should be \textbf{and}] about to get on a zoom call, seated, viewing on a laptop for an hour},'' does not affect the coding of activity type (eating food and screen time). Interestingly, some (9.74\%; 45/462) disagreements in these reports were due to background or irrelevant speech being perhaps correctly captured by ASR but being omitted in the ground truth by our team as they were not intended to be part of the verbal report. For example, this would occur when participants were capturing sedentary activities like \textit{watching TV} and the voice from the TV was also captured.} 

\revised{Even some of the errors involving words that captured activity type could be recoverable. These cases include errors in the verb tenses (\eg, ``\textit{Just came downstairs and \textbf{fix} [should be \textbf{fixed}] me some coffee}...'' [P8]) or compound words (\eg, ``\textit{Walked \textbf{up stairs} [should be \textbf{upstairs}] to second floor}...'' [P6]).
This was also the case for time and effort level. For example, the ASR service often made formatting errors in recognizing time (\eg, ``\textit{Read a book from 6:15 until \textbf{647} [pronounced `six forty-seven'; should be \textbf{6:47}].}''~[P13]), which can be fixed referring to the local context. A disagreement in P6's report, ``...~\textit{Standing, \textbf{minimal} [should be \textbf{minimum}] level of exertion},'' does not affect the coding of effort level category. 

If we had relied solely on the ASR output for their corresponding reports, 82 (out of 651; 12.60\%) would have affected our coding of activity type, time, or effort level. For example, it is challenging to extract time from the ASR output of P11's report, ``\textit{Since about \textbf{132 frozen 245} [should be \textbf{1:30 to present, 2:45}]}...,'' without listening to the audio record. In addition, verbs were sometimes recognized as a totally different one, changing the original meanings in text (\eg, ``\textit{I am just \textbf{resting} [should be \textbf{dressing}] after taking a shower} ...'' [P5]).
We anticipate that automated solutions may be more susceptible to some of these errors.}

\subsection{Participants' Experience with MyMove}
\label{sec:results:experience}



Following the week-long data collection period, we conducted debriefing interviews and guided participants to reflect on their experiences. Their responses helped us understand both strengths and challenges in using MyMove to create verbal activity reports. Participants provided feedback on their experience using MyMove interface and the smartwatch device, specifying information components for reporting, when they used prompted or voluntary methods, and preferences in using virtual vs. physical buttons. \revised{At the end of the debriefing interview, all participants agreed to be contacted for a future follow up session in the project, acknowledging \cameraready{their interest in contributing to this project}.}

\subsubsection{Reactions to the MyMove interface and smartwatch}
\label{sec:results:MyMoveUI}
Participants seemed to have a generally positive experience with MyMove on the smartwatch. 
Ten participants noted that both the interface and the smartwatch contained features that made reporting easy. For instance, P1 commented the flexibility in having multiple reporting methods (``\textit{I think it was easy enough to report, because I was allowed to, you know, report it in various ways}''), and explained physical features of the smartwatch that were favorable (``\textit{the size of the screen is good for my age group, and as well as the buttons were relatively, easily to access}''). P5 mentioned how the multiple modalities helped with the reporting process (``\textit{It was very efficient watch. It was nice that you could just either touch [the screen] or the [physical] buttons.}''). Participants also appreciated the text on the screen, indicating the type of information components to include when recording their activity reports (``\textit{I'd remember what information I had to give you so that was very helpful for me.}'' [P10]).

\revised{On the other hand, participants faced challenges when interacting with the system. 
At the debriefing interview, six participants mentioned that the watch occasionally did not respond to their touch and that they had to click on the Record button a couple of times to start the recording. For example,} P3 mentioned, ``\textit{There were times when I thought I'd recorded something... it seemed like the watch was telling me I hadn't recorded it. So I recorded it again.}'' \revised{We reflect on this challenge and an alternative design in Section  ~\ref{sec:disc:watchUI}.}
Some participants expressed concerns with wearing the smartwatch long-term (``\textit{I don't know if I would want to wear this watch all the time to do it}'' [P6]), and with the smartwatch’s battery life (``\textit{I didn't have any challenges with the watch, except for the fact that it ran out of battery.}'' [P9]).  

\subsubsection{Reactions to specifying information components}


When reflecting on their experience with specifying the activity, timespan, and effort, many participants reacted positively. Several found that describing their activities was relatively easy: (``\textit{It was easier than I thought it would be}'' [P10]), (``\textit{I didn't really have any problems with it. It was pretty straightforward}'' [P7]), (``\textit{If you just wanted what I was doing at the moment, I didn't find that difficult}'' [P2]). 

Participants also expressed challenges in specifying the level of effort required and the time taken.
Seven participants reported having difficulty describing their effort level since it was hard to determine, especially for activities involving multiple tasks (``\textit{In the midst of that activity, I did something else that may have changed the amount of energy required, ...the effort was the hard one for me to actually document that piece of it}'' [P1]). To help determine effort, some participants would use physiological indicators such as breathing, muscle strain, tiredness, or even their exercise performance (``\textit{I can just look at my Strava recording and give you a time and a speed, which sort of gives you an intensity}'' [P4]). 
Six participants found specifying time components to be challenging, including recalling the amount of time taken or just remembering to add time components to the activity description. Eight participants utilized different strategies to assist with tracking activity timespan. Methods included using their memory, a device, or even writing the time down in order to remember the time (``\textit{Whenever I started an activity, I would just look at the watch before I started, and then try to record it right afterwards, so I had it right there}'' [P13]).

\subsubsection{Situations for using voluntary and prompted methods}

In addition, participants described situations in which they would use voluntary and prompted reporting. We found that each method had unique advantages, including having the freedom to report voluntarily at any time and being aware about reporting activities due to prompted notifications. Some participants appreciated the pings because they served as a reminder to record the activity right away or after finishing the activity. P8 commented how ``\textit{it was good to have it, because it reminded you that maybe you hadn't recorded what you were doing,}'' and P1 stated, ``\textit{Had it not been for the watch's alert, I'm not quite sure that I would have captured that information as well.}'' When asked what participants disliked about receiving ping notifications, six participants said that pings were delivered during inopportune moments (``\textit{There were a couple of times when I just couldn't answer the ping. That was in a meeting or something}'' [P7]). Participants also reported that pings seemed unnecessary for redundant activities (``\textit{I was... reporting the same thing all the time}'' [P7]), or too frequent for longer activities which they had already reported earlier (``\textit{I'm just doing the same thing I recorded that I was doing before}'' [P8]).

\subsubsection{Preferences between virtual vs. physical buttons}
Eight participants preferred using virtual buttons, whereas two participants stated a preference in using the physical buttons. Those who preferred using the virtual explained how virtual buttons were more familiar and convenient (``\textit{I'm very used to using touch screens all the time. My instinct was just natural to go there}'' [P5]), and easier to use and understand (``\textit{It was just easier for me to tap to screen}'' [P1]). Some even expressed confusion in how the physical buttons would work (``\textit{I wasn't always sure what [the physical button] was going to do, or... how it was going to respond}'' [P7]). As we mentioned in Section ~\ref{sec:results:MyMoveUI}, some participants had trouble getting the virtual buttons to respond accordingly due to having difficulty with the wake-up functionality associated with using the virtual buttons.

\section{Discussion}

In this section, we first reflect on several aspects that are related to the feasibility of MyMove in facilitating data collection with older adults, such as engaging older adults in activity labeling, using the verbal reports as an information source for activity labeling, and capturing older adults' activities in a comprehensive manner. We also discuss how our findings and the data older adults collected using MyMove can be used toward creating personalized activity trackers that attune to idiosyncratic characteristics of individual users and their unique needs. We then discuss limitations in our study that may affect the generalizability of our findings as well as future work.

\subsection{Engaging Older Adults in Activity Labeling}
\label{sec:disc:engaging}

We were pleasantly surprised by the high adherence and engagement of our participants in the one-week data collection. On average, they wore the smartwatch for 11.6 hours and activPAL for nearly 24 hours every day. 
Furthermore, our participants submitted 13.45 reports per day on average even though the compensation was not tied to the number of reports. Given that the most challenging aspect of an EMA study is its high data capture burden and frequent interruptions~\cite{Berkel2017ESM}, we believe that this is a promising outcome. 
Participants' positive feedback on MyMove indicates that the system itself may have contributed to this high adherence; it provided flexible ways to capture data using speech, including voluntary and prompted reporting methods as well as simplified data capture flow and UI. 
Even though all but one experienced the smartwatch for the first time through our study, all participants could use MyMove without much trouble. 

In debriefing, however, most participants stated that collecting data in this manner would not be sustainable, and the one-week duration would probably be the maximum they could continuously engage at this level. 
As is common with other ESM studies, our study imposed a high burden on the participants, for example, having them consciously think of activity type, start/end time, and effort level when they receive hourly notifications. 
In our study, we did not limit the scope of what activities are report-worthy because our goal was to examine the feasibility of collecting in-situ activity labels with older adults using speech on a smartwatch.
Going forward, such comprehensive data capture may not be necessary; we expect that the burden of capturing activity labels would be reduced when we have a fixed set of targeted activities (\eg, walk, gardening, golf, yoga) that require labeling and better mechanisms for estimating activity timespan (\eg, by automatically detecting abrupt changes in sensor data) \revised{and effort level (\eg, by leveraging heart rate data)}.

\subsection{Leveraging Verbal Reports as an Information Source for Activity Labeling}


It was encouraging to see that all of the 1,224 verbal reports are valid and that researchers could transcribe and understand all of them. Although some participants, in the debriefing interview, mentioned that they accidentally triggered the recording, it seemed that they were able to cancel it or the recording was timed out and thus erased.
This demonstrates that, despite the challenges older adult participants faced with the unfamiliar technologies, they still could successfully submit valid reports using our novel data collection approach.  
Furthermore, the word error rates by two state-of-the-art automatic speech recognition systems were relatively low: 4.93\% with Microsoft Cognitive Speech and 8.50\% with Google Cloud Speech. We were reassured that Microsoft Cognitive Speech's error rate on our older adult participants is lower than what Microsoft had reported in 2018 (5.0\%). \revised{Nonetheless, these numbers serve merely as anecdotal evidence. Our small sample is not representative of older adults; all participants were native English speakers in the US and none of them identified as disabled. Of course, speech as an input modality, can be advantageous for many disabled people such as blind individuals~\cite{azeknot2013exploring, ye2014current, hong2020reviewing} and those with upper limb motor impairments~\cite{hup2012dual, malu2018exploring}. However, it is still limited for dysarthric~\cite{malu2018exploring, derussis2019impact}, deaf~\cite{glasser2017feasibility}, and accented~\cite{prasad2020accents} speech as well as for low resource languages and noisy environments, all being active areas of research. With advances in speech recognition, we believe that} we could leverage \revised{for many} older adults \revised{their} verbal reports as a reliable data source in an automated manner. This opens up an opportunity for automatically extracting user-generated activity labels (type and semantics).

\revised{That said, inferring quantitative or ordinal data from free-form text is not a trivial problem, as in the case of the effort level coding. As such, data such as effort level would be better off if they are collected in a structured way (\eg, have people select from a scale or predefined categories). This would require new UI \& interaction designs \eg, leveraging a simple touch interaction on a smartwatch or predefined voice commands.} 
\subsection{Comprehensive Capturing of Older Adults' Activities}

Even though understanding daily activities of older adults was not the main goal of our study, the collected data seem to cover more classical types of activities older adults perform while reflecting recent trends. We consider this as additional evidence to demonstrate the feasibility of in-situ data collection with older adults.

The types of activities emerging from our participants' reports overlap with most of the activities reported in prior literature, though their naming/grouping may not be fully aligned. For example, an interview study with 516 German older adults conducted in 1996 identified 44 types of activities and grouped them into eight categories---\textit{Personal Maintenance}, \textit{Instrumental ADLs}, \textit{Reading}, \textit{Television}, \textit{Other Leisure}, \textit{Social Activities}, \textit{Paid Work}, and \textit{Resting}~\cite{Horgas1998OlderAdultsLife}. While these activities appeared in our dataset, we categorized them differently, for example, the Reading activity type under Hobby/leisure and many of the Instrumental ADL activities under Housekeeping. We note that there is not a clear consensus on how to group activities, so it will be important to preserve both raw labels and coded categories for future reference. 

New activities have also emerged as digital technologies have advanced over time.
For example, our Screen time category includes ``Computer,'' ``Mobile device,'' and ``TV,'' whereas prior studies conducted in 1982~\cite{Moss1982TimeBudgetOlder} and 1998~\cite{Horgas1998OlderAdultsLife} had the ``Watching TV'' as the top-level category (without the notion of screen time).
Screen time has absorbed other activities that would have been categorized differently in the past. For example, many of our participants read news on the internet, rather than a printed paper. Screen time also commonly appeared with other activity types, such as social activity (\eg, a video call) and exercise (\eg, online yoga session)\revised{, which may have transitioned from in person to remote during the pandemic}. Labeling systems are bound to evolve as technology changes the way people achieve different functions. Reflecting these changes in designing activity labeling systems, there may be value in having multiple dimensions, such as device type, posture, semantics, to better characterize the captured activities.

\subsection{Capturing Non-Exercise Stepping as a Meaningful Activity for Older Adults}

Researchers have advocated the importance of promoting non-exercise physical activities (free living activities that involve light and moderate physical activities, such as gardening, laundry, cleaning, or casual walking) for older adults~\cite{Sallis2000AssessmentPA, Sparling2015Recommendations, Smith2015PotentialNonExercise}. 
Recent evidence shows that non-exercise physical activities are positively associated with longevity~\cite{Hamer2014NEPASurvival, Ekblom-Bak2014NEPACardio} and cardiovascular health~\cite{Ekblom-Bak2014NEPACardio}, suggesting that \textit{any} activity is better than \textit{no} activity.
However, it is challenging to capture non-exercise physical activities using recall-based methods or sensor-only methods~\cite{Smith2015PotentialNonExercise}. For example, the interview study by Horgas and colleagues identified a generic ``Walking'' category~\cite{Horgas1998OlderAdultsLife}.
Studies that leveraged accelerometer sensors often use statistically determined thresholds for sensor movements and bout duration to categorize activities by intensity level, treating sporadic and light activities as non-exercise (\eg,~\cite{Lee2014UsingAccelerometers, Buman2010LightIntensityOlderAdults}). However, applying uniform thresholds may ignore individualized characteristics~\cite{Schrack2016AssessingPASensors, Rejeski2016LIFE} and there are no standardized thresholds validated for older adults~\cite{Gorman2014AccelerometryOlderAdults}. In addition, most of such studies do not capture user-generated context, potentially important details to understand what people did.

In our study, 14\% of verbal reports (171/1224) contained \textit{non-exercise stepping} with context. We assert that our in-situ data collection method enabled participants to capture more subtle, light-intensity lifestyle activities (\eg, walking around home) that would have not been captured otherwise. 
Older adults' non-exercise stepping activities, which tend to be in slow gaits, are difficult to detect accurately with current waist and wrist-worn accelerometers. 
The context information captured in verbal reports may be a valuable supplement for device-based monitoring to support the training of person-specific classifiers for non-exercise physical activities. 


\subsection{Personalizing Activity Trackers}

Even though our participant sample was relatively homogeneous and geographically constrained to the same area, we observed high variation in the types of activities captured depending both on the participant and on other factors, such as the time of the year. We identified many implications from our findings for the design of personalized activity tracking systems with older adults. 
When automating the tracking of these activities from the sensor data, researchers can potentially leverage some of the existing datasets from younger adults (\eg, WISDM~\cite{weiss2019wisdm,weiss2019smartphone}, UCI-HHAR~\cite{stisen2015smart}, and ExtraSensory~\cite{Vaizman2018ExtraSensoryApp}) to pre-train models for higher level activities, such as sitting, standing, and walking that tend to be common among people regardless of their age. However, preliminary results indicate that model adaptation is necessary as models trained on younger adults' sensor data tend to perform worse on older adults~\cite{fatima2021activity}.
In addition, the diversity in the activity types and semantics among the older adults in our study calls for model personalization beyond adaptation, as one-to-one mapping between older adults' activities and those available in the datasets (from younger adults) may not be possible. We could employ novel model personalization methods like teachable machines~\cite{kacorri2017teachable, hong2020crowdsourcing, lee2019revisiting, dwivedi2021exploring}, which leverage advances in transfer learning~\cite{pan2010survey} and meta learning~\cite{finn2017model,li2021meta-har}.
Systems like MyMove could play a critical role in facilitating this personalization process by supporting older adults and other underrepresented populations in fine-tuning the models in activity tracking applications with their own data, so that the applications can reflect their idiosyncratic characteristics.



\subsection{\revised{Usability Challenges with Smartwatch's Low-power Mode}}
\label{sec:disc:watchUI}
\revised{While our participants had generally positive experiences with MyMove and the smartwatch, six participants occasionally experienced the watch being unresponsive (Section~\ref{sec:results:MyMoveUI}). We suspect that this was caused during Wear OS’s low-power mode (also known as \textit{ambient mode}). When a user is not interacting with their watch for a while, Wear OS automatically enters into the low-power mode, dimming the watch display to save the battery. In this low-power mode, MyMove’s button icons and labels were still visible in low contrast. To make the virtual buttons interactive, participants needed to ``wake up'' the screen by tapping anywhere on the watch display. Alternatively, they could push the physical button to start the recording without needing to wake up the screen.} 

\revised{We showed the buttons and icons in low contrast during the lower-power mode because we wanted to use them as a visual reminder to encourage data capture. However, the low-power mode was, in hindsight, an unfamiliar concept to participants, especially those who are new to a smartwatch: some participants thought that they could interact with the visible button in the low-power mode. Instead of showing the button icons and labels in low contrast, hiding them completely might have been a better design to avoid confusion, which is an interesting design tradeoff we learned from this study.}


\subsection{Limitations and Future Work}
In this section, we discuss the limitations of our study that could impact the generalizability of our findings. 
Although we aimed to recruit participants with diverse backgrounds, our participants are not representative samples of older adults. They were all highly educated (\eg, having a college degree or above), had high-baseline technical proficiency (\eg, being able to use a Zoom video call), and did not have speech, hearing, motor, movement, or cognitive impairments. 
While this work is just a first step toward designing and developing inclusive activity tracking systems, we believe it is important to conduct a follow-up study with older adults with different educational backgrounds, health conditions, and technical proficiencies. This would help us extend our understanding of the strengths and limitations of in-situ data collection with speech on a smartwatch. \revised{As discussed above, we anticipate that this modality can be advantageous for people that were not captured by our small sample such as those who are blind or have low vision, as speech input can be more efficient for this user group~\cite{azeknot2013exploring, ye2014current}. However, it may not be inclusive of dysarthric, deaf, and accented speech, especially if the goal is automatic extraction of the activity labels. Recent speech recognition personalization efforts like Google's Project Euphonia~\cite{macdonald21disordered, green21automatic} are promising. Similarly, efforts that attempt cross-lingual knowledge transfer in speech recognition from high- to low-resource languages (\eg,~\cite{syed2017active, khare21low}) can make speech input more inclusive.  Even then, the challenge of automatically extracting activity labels, timing, and effort levels from verbal reports remains. Information extraction from unstructured reports is an active area of research in natural language processing (\eg, processing medical verbal or written reports~\cite{mustafa2021automated, kersloot2020natural}). Similar to the healthcare context, we could leverage transfer-  and meta-learning techniques to deal with the lack of training data. More so, in contrast to healthcare, we also have an opportunity to shape (\ie, via design and personalization) the user interactions with the activity trackers. Thus, we can influence the structure and vocabulary in the reports to meet the algorithmic capabilities halfway \eg, by optimizing across flexibility, efficiency, and effectiveness for both users and algorithms.}

Our study preparation (\eg, dropping off \& picking up study equipment) and study design provided more face-to-face time with the participants than a typical remote deployment study. This provided a chance for older adult participants to ask questions and troubleshoot issues. Thus, these repeated interactions may have contributed to forming rapport between participants and researchers, which in turn, could have contributed to the high engagement. We had two onboarding sessions with the 4-day adaptation period in between. During the 4-day adaptation period, participants became used to wearing and maintaining (\eg, charging) the devices, and were ready to collect data on Day 5 of the study. Some participants explicitly mentioned that the tutorial and the adaptation period were critical for their engagement. For example, P5 commented, ``\textit{Giving me a few days to get used to the equipment and how it worked... Making sure I had plugged in and make sure that I was charging, how to record, and I thought that was really good. And just how to give the reports, I think the orientation was very helpful as well.}'' While we believe that giving a good tutorial before the actual experiment is important, we acknowledge that our particular approach may not scale. \cameraready{In addition, the study compensation and participants' interest in contributing to a research project may also have affected participants' engagement, although it is common to incentivise participants in ESM studies.}

We note that we did not collect information on medication use of participants. Medications, such as $\beta$-blockers, can influence heart rate and may blunt the response to higher intensity exercise, resulting in lower heart rate measurements. As such, the intensities we recorded during \textit{Strenuous} activities may be not accurately reflect the degree of vigor with which the participant was being active, resulting in the percentage of HR$_{max}$ that were closer to the low intensity activities. Future work should consider the incorporation of participants' medication information to further validate heart rate intensities, especially for high-intensity activities. 

We chose a smartwatch as an only means to collect verbal activity reports and to deliver notifications. In the future, we can leverage other ``smart'' devices for more comprehensive and accurate data collection. For example, when the TV is on and the person is nearby (without much movement), we can infer that the person is watching TV.
In addition, we can leverage the speech input capability of other devices. For example, a person can report their activities using a smart speaker that is becoming more prevalent (the speaker can even play the recording back to the person). Similarly, a person can record their activities from their smartphone, tablet, laptop, or desktop as all these devices are equipped with a microphone. Since these devices have larger display than a smartwatch, people can view or edit data they captured elsewhere (\eg, a smartwatch or smart speaker) from these devices.

In our study, we did not provide feedback other than the number of reports participants submitted on a given day. However, in the debriefing interview, half of our participants reported that they became more aware of the activities they performed and how they spent the time. Also known as the ``reactivity effect,'' this is a well-known phenomenon in behavioral psychology~\cite{Nelson1981Reactivity}. 
We believe that our approach to collecting in-situ data can serve a dual purpose of \textit{activity labeling} and \textit{self-monitoring}; the latter can be further augmented through providing informative and engaging feedback---for example, showing how much they have been sitting, working out, gardening---and people may be more motivated to engage in desirable activities while capturing data (labels). 

\section{Conclusion}

In this work, we examined the feasibility of collecting in-situ activity reports with older adults, with the ultimate goal of developing personalized activity tracking technologies that better match their preferences and patterns.
We built MyMove, an Android Wear reporting app. Considering older adults as the main user group, we streamlined the data capture flow and leveraged the flexible speech input on a smartwatch.
Through a 7-day deployment study with 13 older adults, we collected a rich dataset including older adult participants' verbal reports, the sensor data from a smartwatch and a thigh-worn activity monitor, and participants' feedback from the debriefing interviews.
Our results showed that participants were highly engaged in the data collection. They submitted a total of 1,224 verbal reports. Additionally, the wear time of the smartwatch (11.6 hours/day) and thigh-worn activity monitor (23.3 hours/day) was very high.
Examining the verbal reports further, we found that all of them were valid, that is, a researcher could understand and transcribe them. Moreover, verbal reports could be transcribed with state-of-the-art automatic speech recognition systems with acceptable error rates (\eg, 4.93\% with Microsoft Cognitive Speech).
These results, taken together, indicate that our novel data collection approach, realized in MyMove, can facilitate older adults to collect useful in-situ activity labels.
Going forward, we are excited to continue our endeavors towards building personalized activity tracking technologies that further capture meaningful activities for older adults.


\begin{acks}
We thank our study participants for their time, efforts, and feedback. We also thank Catherine Plaisant, Yuhan Luo, and Rachael Zehrung for helping us improve the tutorial protocol. We are also grateful for Bonnie McClellan and Explorations On Aging for helping us recruit study participants. This work was supported by National Science Foundation awards \#1955568 and \#1955590.
\end{acks}

\balance
\bibliographystyle{ACM-Reference-Format}
\bibliography{bibliography}

\end{document}